\documentstyle [preprint,aps]{revtex}

\begin{document}

\begin{title} {The Electron-Phonon Interaction in the Presence
of Strong Correlations }
\end{title}

\author{M. Grilli and C. Castellani}
\address{Dipartimento di Fisica, Universit\'a La Sapienza,
Piazzale Aldo Moro, 00185 Roma, Italy}
\maketitle

\begin{abstract}
We investigate the effect of strong electron-electron
repulsion on the
electron-phonon interaction from a Fermi-liquid point of view.
In particular we show that the strong interaction is responsible
for vertex corrections, which are strongly dependent on the
$v_Fq/\omega$ ratio, where $v_F$ is the Fermi velocity and $q$ and
$\omega$ are the transferred momentum and frequency respectively.
These corrections generically lead to  a strong suppression of the
effective coupling between quasiparticles mediated by a single
phonon exchange in the
$v_Fq/\omega \gg 1$ limit. However, such effect
is not present when $v_Fq/\omega \ll 1$. Analyzing the Landau
stability criterion, which involves the effective interactions in
the dynamical limit, we show that a sizable electron-phonon
interaction can push the system towards a phase separation
instability.

A detailed analysis is then carried out using a slave-boson
approach for the infinite-U three-band Hubbard model describing the
basic structure
of a ${\rm{CuO_2}}$ plane in copper oxides. In the presence
of a coupling between the local hole density and a dispersionless
optical phonon, we explicitly confirm the strong dependence of the
hole-phonon coupling on the transferred momentum versus frequency
ratio. We also find that the exchange of phonons
leads to an unstable phase with negative compressibility   already
at small values of the bare hole-phonon coupling. Close to the
unstable region, we detect Cooper instabilities both in s- and
d-wave channels  supporting a possible connection between phase
separation and superconductivity in strongly correlated systems.
\end{abstract}


\narrowtext


\section{Introduction}
The problem of the interplay between the electron-phonon (e-ph)
coupling and the (strong) electron-electron (e-e)
interaction  is an interesting problem, which still lacks a
complete understanding. At present this topic is very hot
since  various facts indicate that  the lattice can play a non
negligible role in  both the superconducting copper oxides and the
fullerenes.  As far as these latter materials are concerned,  the
strength of the interaction is still a matter of a debate,  but
there is a rather large agreement on the prominent role of the
lattice.  On the other hand, as far as copper oxides are concerned,
it is generally  recognized that the e-e interaction is very large
in these systems, but  it is the relevance of the lattice that is
 questioned. However various groups
\cite{stanford}-\cite{ranninger} claim that there are
both experimental evidences
and theoretical arguments supporting a prominent role of the
lattice in the cuprates. Moreover recent optical experiments in the
mid-infrared frequency region indicate the presence of polaronic
effects \cite{polaronexper} for the very lightly doped compounds,
which are known to be strongly correlated systems.

Various issues can be addressed in investigating
the role of the lattice in the presence of a strong e-e
interaction. In particular two questions can be raised, which are
relevant, both on a general ground  as well as in the framework of
high temperature superconductivity. The first question concerns
the possibility of having a large phonon-mediated effective e-e
coupling due to the strong mass enhancement ($m^*/m \gg 1$)
occurring in a strongly correlated Fermi liquid: after all
the dimensionless coupling $\lambda=\gamma^2\nu_0$
($\gamma$ is the usual bare electron-phonon coupling and $\nu_0$ is
the free-electron density of states) \cite{AGD} could
grow very large because of the density of states
renormalization arising from the
mass enhancement $\lambda \to \lambda=\gamma^2\nu^*$, with
$\nu^*=\left(m^*/m\right)\nu_0$.
This would have crucial consequences  both on transport properties
as well as on the Cooper pair formation. Moreover, a large
$\lambda$ would favor the formation of polarons.

The second question regards the possible occurrence of
instabilities in the electronic gas. Previous studies of single- and
multi-band Hubbard models in the strong-coupling ($U \gg t$) limit
have revealed a strong tendency of these systems to undergo Phase
Separation (PS) and Charge-Density-Wave (CDW) instabilities as soon
as short range interactions are introduced. This occurs
irrespectively to the magnetic (e.g. nearest neighbor Heisenberg
coupling) \cite{marder}-\cite{dagotto} or the Coulombic (e.g.
nearest neighbor  repulsion) \cite{GRCDK1}-\cite{RCGBK} nature of
the short-range interaction.
In the framework of interest here one can ask
whether also an e-ph coupling can destabilize the electron gas.
The analysis of the stability with respect to PS of a given model
is particularly relevant since the investigation
carried out in models displaying PS also showed that
superconductivity takes place close to the instability region
\cite{GK}, \cite{GCK},\cite{dagotto}-\cite{GRCDK2}
 (the Cooper pairing occurring as a precursor of PS due to the
attraction eventually driving the system to the PS instability).

We anticipate here the answers to the two above questions, which
will be analyzed in detail in the rest of the paper.

As far as the first problem is concerned, a general analysis
performed within a standard  Fermi-liquid scheme
with phonons coupled with the electron density reveals a
dependence of the effective  phonon-mediated e-e interaction on the
ratio between
the transferred momentum $q$ and the frequency $\omega$,
because of the vertex corrections generated by the e-e interaction.
This dependence can be  particularly strong in the presence of a
large quasiparticle mass  enhancement.
As a result, when $v_F q/\omega
\gg 1$, we find that the screening of the quasiparticles
strongly enhances the vertex corrections. These corrections tend to
suppress the effective interaction so that the resulting static
coupling $\lambda$ is small.
This is in agreement with previous
calculations \cite{kim} performed in a specific model, where,
however, the e-ph coupling arises from the ion-position dependence
of the hopping matrix elements (the so-called covalent e-ph coupling).
On the contrary, in the opposite limit $v_F q/\omega \ll 1$
the intraband quasiparticle screening is ineffective, the
physics being dominated by high energy (e.g. interband
charge-transfer  (CT)) processes, and a large effective e-ph
interaction results.

The strong dependence of the vertex corrections on the $v_F q/\omega$
ratio renders the analysis of the effects of the  e-ph coupling
particularly delicate, since different physical  quantities may
involve different $v_F q/\omega$ regimes. In particular the
e-ph scattering in transport properties is
dominated by low-energy-high-momentum processes, which suffer a strong
suppression due to the large vertex corrections.
As pointed out in Ref.\cite{kim}, this
will reduce the relevance of the e-ph scattering particularly at low
doping, where the e-e interaction effects are more relevant.
Similar conclusions have also
been recently drawn \cite{zeyher} from the
analysis of a single-band Hubbard model with electrons coupled with
an optical dispersionless phonon.

On the other hand an affirmative answer can be given to the
second question concerning the occurrence of instabilities in
the electron gas. In fact, while the behaviour of $\lambda$
described above is generally true far from an instability, it will
be shown that a sufficiently large e-ph coupling can instead
produce a PS instability \cite{notamax}. This
reflects the fact that the Landau
stability criterion for the symmetric Landau parameter,
$F_0^s>-1$, required for a positive compressibility, involves the
total (e-e and e-ph mediated) interaction in the dynamical limit
$v_Fq/\omega \to 0$. At the point where
$F_0^s=-1$ the compressibility
diverges and nearby, even the total static scattering
amplitude (within a RPA resummation of the phonon-mediated
interaction) is large and negative. Also this analysis can be
carried out on a rather general ground. However, the quantitative
determination of the instability conditions, being  related to the
subtle interplay between various interactions, depends on the
couplings involved and must rely on the treatment of a specific
model. For this purpose we shall consider a three-band Hubbard model
describing the holes in the ${\rm{CuO_2}}$ planes of
the high ${\rm{T_c}}$ copper
oxides. A coupling between the local hole density and a
dispersionless optical phonon will be specifically considered. To
deal with the  electronic correlations in the strong-coupling limit
we use the standard slave-boson technique  within a
$1/N$ expansion. The analysis of this model  explicitly detects the
presence of an instability region where the compressibility of the
fermion gas diverges and then becomes negative. Near the region
where the system becomes unstable, superconducting instabilities
are found by averaging over the Fermi surface the
interaction amplitudes in the Cooper channel.

The plan of the paper is as follows. In Section II we address the
two above questions working in the general framework of the Landau
Fermi-liquid theory. The three-band Hubbard model with the electrons
coupled with a dispersionless optical phonon is introduced in Section
III. The instabilities are investigated in
Section IV. Final remarks are contained in Section V.

\section{Phonon-mediated effective interaction: a Fermi-liquid
discussion}
One of the basic concepts of Landau Fermi-liquid theory
is the idea of mass renormalization. This idea is also crucially
present in the most common treatments of the interacting Fermi
systems. In particular it is naturally introduced in the Gutzwiller
treatment of the Hubbard model \cite{gutzwiller} and is at the
basis of the Mott-Hubbard transition in this model
\cite{brinkmanrice}. This very same basic concept recurs in other
popular techniques like, e.g. the slave-boson technique
\cite{barnes}-\cite{slabos}.

In the presence of a strong interaction the mass renormalization
can be very large,  $m^*/m \gg 1$,  and results in an enhancement
of the quasiparticle density of states
$\nu^*=\left(m^*/m\right)\nu_0$. Then, as already discussed in the
previous Section, the natural question on the
consequences of this renormalization on the effective phonon
mediated e-e interaction is whether the
bare (free electron) effective e-e coupling $\lambda=\gamma^2\nu_0$
grows into $\lambda=\gamma^2\nu^*$ as an  effect of the mass
enhancement.

Without a significant loss of generality we address the above
question discussing the case of an optical phonon coupled to the
local density of electrons by a constant coupling $\gamma$
\cite{notagam}.
In the presence of a (possibly large) e-e interaction one has to
worry about the e-ph vertex corrections involving the e-e
interaction, for which no Migdal theorem can be applied.
The problem is conveniently cast in the language of the
standard Fermi-liquid theory \cite{nozieres}, by using the two
relations connecting the density vertex  $\Lambda(q,\omega)$ and
the wavefunction renormalization $z_w$ in the dynamic and static
limits
\begin{eqnarray}
z_w\Lambda\left(q=0,\omega \to 0\right) & = & 1  \label{wi1}\\
z_w\Lambda\left(q\to 0,\omega = 0\right) & = &
{{1}\over{1+F_0^s}}\,\, . \label{wi2}
\end{eqnarray}
$F_0^s=2\nu^*\Gamma_\omega$  is the standard Landau parameter
and $\Gamma_\omega$ is the
dynamic ($q=0$, $\omega \to 0$) effective e-e scattering amplitude
between the quasiparticles.

We first consider the effective e-e interaction
arising from a single phonon exchange at lowest order in
$\gamma^2$. Then the vertex corrections
can not include phonon processes and the dynamic Landau scattering
amplitude in Eq.(\ref{wi2})
is due to the e-e interaction only. To explicitly keep
memory of this limitation we append a suffix ``$e$" to
$\Gamma_\omega$ and to any quantity not involving phononic processes.
Thus $F_0^{s(e)}=2\nu^*\Gamma_\omega^e$.
The relations (\ref{wi1}) and (\ref{wi2}) are exact Ward
identities, which must be satisfied irrespective of the details of
the e-e interaction and show a drastic
difference between the dynamic $(q=0, \omega\to 0)$ and static
$(q\to 0,\omega= 0)$ limits. Whenever the exchange of  a
phonon  takes place, the vertex
corrections must be included leading to a different behaviour of
the effective interaction in the two limits. The  effective
dimensionless e-e interaction mediated by a single phonon exchange
reads
\begin{equation}
\nu^*\Gamma_{\rm{eff}}^{ph}
(q,\omega)=\nu^*\gamma^2{z_w^{e}}^2{\Lambda^{e}}^2(q,\omega)
{{\omega^2(q)}\over{\omega^2-\omega^2(q)}} \label{phoneffint}
\end{equation}
where  $\omega(q)$ is the phonon dispersion.
Here the presence of $z_w^e$ indicates that we are considering
the effective interaction between quasiparticles and $\Lambda$
expresses the difference of the phonon coupling to the
quasiparticles with respect to particles \cite{nozieres}.

The effects of the
strong $\omega-q$ dependence of the electronic density vertex
$\Lambda$,  in $\Gamma_{\rm{eff}} (q,\omega)$ (see Eq.
(\ref{phoneffint})) can be made apparent in the small q and $\omega$
limits, where the relations (\ref{wi1}) and (\ref{wi2}) can be
used. Then one obtains
\begin{eqnarray}
\nu^*\Gamma_{\rm{eff}}^{ph} \left(q\to 0,\omega \to 0\right) & = &
 -\gamma^2\nu^*, \,\,\,\,\,\,\,\,\,\,\,\, {{v_Fq}\over{\omega}} \to 0
\label{1phdyn}\\
\nu^*\Gamma_{\rm{eff}}^{ph} \left(q\to 0,\omega \to
0\right) & = &-\gamma^2  {{\nu^*}\over{\left(1+F_0^s\right)^2}}
 =\nonumber \\
& =& -\gamma^2  {{\kappa^e}\over{\left(1+F_0^{s(e)}\right)}}
\nonumber \\
& = & -\gamma^2 {{\kappa^e}^2 \over \nu^*}\,;
\,\, {{\omega}\over{v_Fq}} \to 0
\label{1phstat}
\end{eqnarray}
where $\kappa^e ={{\nu^*}\over{\left(1+F_0^{s(e)}\right)}}$ is the
compressibility of the Fermi liquid in the absence of coupling
with the lattice. It should be noted that, in the present case  of
an optical phonon, the phonon propagator tends to -1 in both
limits, Eqs.(\ref{1phdyn}) and (\ref{1phstat}).

The difference between the dynamic and static case can be
dramatic in the case of a Fermi liquid
with a large mass enhancement  $m^*/m \gg 1$,
but with a negligible compressibility renormalization
($\kappa^e \simeq \nu_0$) \cite{notafl}. In fact, in the case
under consideration,
$F_0^{s(e)}$ is  proportional to the
quasiparticle density of states, $\nu^*=\left(m^*/m\right)\nu_0\gg
\nu_0$,  and one has $F_0^{s(e)} \gg 1$.
Then  Eqs.(\ref{1phdyn}) and (\ref{1phstat}) read
\begin{eqnarray}
\nu^*\Gamma_{\rm{eff}}^{ph} \left(q= 0,\omega \to 0\right)
&=&-\gamma^2  \left({{m^*}\over{m}}\right)\nu_0 \label{1} \\
\nu^*\Gamma_{\rm{eff}}^{ph}
\left(q\to 0,\omega = 0\right) &= & -\gamma^2 {{\kappa^e}^2 \over
\nu^*} \nonumber \\
& \simeq & -\gamma^2
\left({{m}\over{m^*}}\right)\nu_0  \label{2}
\end{eqnarray}
so that the effective one-phonon-mediated e-e interaction is large
($\sim m^*/m$) in the dynamic limit and small ($\sim m/m^*$) in the
static one.

The strong $\omega-q$ dependence  in Eqs.(\ref{1}) and (\ref{2})
concerns the small-$q$ and small-$\omega$ limits. This result
relies on quite general arguments, whereas the case of finite
$q$'s and $\omega$'s will need the analysis of a specific model
and will be discussed in the context of the three-band Hubbard
model in the next sections. Our expectation, which will be
confirmed by the analysis in Section IV, is that the product
$z_w\Lambda$ will be roughly of order one (dynamical limit)
all over the region outside the particle-hole continuum, while it
will strongly deviate from unity in the region of the
particle-hole continuum, where screening processes take place.
As a consequence the e-ph coupling (and the e-e interaction
mediated by phonons) will be depressed by the e-e interaction
in all processes involving small energy and large momenta (as in
the low energy lifetime and transport).

The above conclusion on the irrelevance of the e-ph coupling is
based on a lower order analysis in $\gamma^2$.  It evidently
contrasts with the fact that the Landau stability criterion
$F_0^s>-1$ involves the full e-e interaction in the dynamical
limit, where the e-ph coupling is not depressed by the ``pure''
e-e vertex corrections. Within the same limits discussed above
(lowest order in $\gamma^2$) we have
\begin{equation}
F_0^s=\nu^*\left( \Gamma_\omega^e-\gamma^2 \right) \label{7bis}
\end{equation}
indicating that a sizable $\gamma^2$ can indeed lead to
$F_0^s<-1$. (Note that $m^*/m \gg 1$ requires a large
bare repulsion in units of the bare
Fermi energy, as in the single- or multi-band Hubbard model near
the metal-insulator transition, but this does not imply a large
$\Gamma_\omega^e$). The lowest order analysis showing the
depression of the e-ph coupling in the low energy processes
maintains its full validity with respect to the inclusion of
higher order terms provided $F_0^s$ in Eq.(\ref{7bis}) is still
of order $m^*/m$. On the contrary, near the
instability condition $F_0^s=-1$,
the phonon contributions to the vertex cannot be neglected
and the e-ph interaction is relevant even in the static
limit. To clarify
this point we extend our analysis considering a bubble
resummation of the one-phonon processes, in the framework of a
standard RPA approach for the quasiparticles.

As a first step we introduce an effective dynamic e-e scattering
amplitude $\Gamma_{\omega}^{e} (q,\omega)$ {\it{with no phonon
processes included in it}}. Here and below, the subscript
$\omega$ indicates that {\it{no intraband screening processes}} are
present in the considered quantity. The $q\to 0$ and
$\omega \to 0$
limit of  $\Gamma_{\omega}^{e}(q,\omega)$ reproduces the
above Landau dynamic scattering amplitude $\Gamma_\omega^e$ of the
Fermi liquid in the absence of e-ph coupling.

The second ingredient of our analysis is the
intraband Lindhardt polarization bubble
\begin{equation}
\Pi(q,\omega)=2\sum_k {{f\left( E_{k+{q \over 2}}\right)-
f\left( E_{k-{q \over 2}}\right)} \over
{E_{k+{q \over 2}}-E_{k-{q \over 2}}- \omega}}
\end{equation}
where $f\left( E_{k}\right)$ is the Fermi function and $E_k$ is
the quasiparticle band.
Here we have included the spin multiplicity in the
definition of $\Pi (q,\omega)$. This bubble can be dressed with all
possible intraband and interband purely electronic screening
processes within a RPA resummation leading to
\begin{equation}
{\widetilde {\Pi}}^e(q,\omega)={{\Pi(q,\omega)}\over
{1+\Gamma_{\omega}^{e} (q,\omega)\Pi(q,\omega)}}
\end{equation}
Finally we consider the self-energy corrections to the phonon
propagator arising from the purely electronic processes.
Once both the interband and the intraband  screening of
the quasiparticles is taken into account one obtains
a phonon propagator with a self-energy correction given by
$g^2{\widetilde {\Pi}}$
\begin{eqnarray}
{\cal D}(q,\omega) & = & {{\omega^2(q)}\over{\omega^2-\omega^2(q)
\left(1-\gamma^2{\widetilde \Pi}^e(q,\omega)\right)}} \nonumber \\
& = & {{\omega^2(q)}\over{\omega^2-\omega^2(q)
{{1+{\widetilde \Gamma}_\omega (q,\omega)\Pi(q,\omega)}\over
{1+\Gamma_{\omega}^{e} (q,\omega)\Pi(q,\omega)}}
}}. \label{phonprop}
\end{eqnarray}
Notice that the quantity
\begin{equation}
{\widetilde \Gamma}_\omega(q,\omega)\equiv
\Gamma_{\omega}^{e} (q,\omega) +\Gamma_\omega^{ph}(q=0,\omega \to 0)=
\Gamma_{\omega}^{e} (q,\omega) -\gamma^2 \label{gamtil}
\end{equation}
has been introduced. Taking the limit $q \to 0$ first and then
$\omega
\to 0$, ${\widetilde \Gamma}$ becomes the Landau dynamic scattering
amplitude {\it{in the presence}} of an e-ph coupling.

We now can evaluate the total scattering amplitude splitting the
contribution of the purely electronic processes and the
contribution of the processes also involving phonons
\begin{equation}
\nu^*\Gamma (q,\omega)= {{\nu^* \Gamma_{\omega}^{e} (q,\omega)} \over
{1+\Gamma_{\omega}^{e} (q,\omega)\Pi(q,\omega)}} +
{{1} \over {\left[
1+\Gamma_{\omega}^{e}(q,\omega)\Pi(q,\omega)\right]^2}}
{{\nu^*\omega^2(q) \gamma^2}
\over{\omega^2-\omega^2(q)
{{1+{\widetilde \Gamma}_\omega (q,\omega)\Pi (q,\omega)} \over
{1+\Gamma_{\omega}^{e} (q,\omega)\Pi(q,\omega)}}
}} \label{totgam}
\end{equation}
It is important to notice that vertex corrections have been
included in the second term by specializing Eqs.(\ref{wi1}) and
(\ref{wi2}) into $z_w^e\Lambda^e=
\left[ 1+\Gamma_{\omega}^{e}(q,\omega)\Pi(q,\omega)\right]^{-1}$.
Nevertheless,
now the attraction mediated by this second term can be large even
in the static limit if the denominator of the phonon propagator
becomes small. In fact, in the $\omega \to 0$ limit one has for the
attractive part of Eq. (\ref{totgam})
\begin{equation}
-{{1} \over
{ \left[ 1+\Gamma_{\omega}^{e}(q,\omega)\Pi(q,\omega)\right]}}
{{\nu^* \gamma^2}
\over{1+{\widetilde \Gamma}_\omega (q,0)\Pi(q,0)}}.\label{statattr}
\end{equation}
{}From this expression one can see that
the condition for a diverging  static scattering amplitude,
eventually leading to an instability of the Fermi liquid, results
in the  condition on the Landau dynamic scattering amplitude
\begin{equation}
1+{\widetilde \Gamma}_\omega
(q \to 0,\omega=0)\Pi(q \to 0, \omega=0) =0
\label{instcond}
\end{equation}
equivalent to the usual $1+F_0^s=0$, Eq.(\ref{7bis}), since  in
this limit $\Pi(q,\omega)$ reduces to $2\nu^*$.

This result  shows that a Fermi liquid can indeed be
destabilized by the coupling of the quasiparticles to the
lattice even in the presence of a very strong e-e repulsion.
It must be noticed that this instability does not imply
the full phonon softening, since the determination of the
renormalized phonon frequency ${\widetilde \omega}(q)$ requires the
value of
\begin{equation}
{{1+{\widetilde \Gamma}_\omega (q,\omega)\Pi (q,\omega)}\over
{1+\Gamma_{\omega}^{e} (q,\omega)\Pi(q,\omega)}}
\end{equation}
in a range of frequencies, where the numerator does not vanish.
The instability appears instead as an overdamping of the zero
sound driven by the e-ph mediated attraction.

The quantitative determination
of the needed strength of the e-ph coupling in order to have an
instability  must rely on the  analysis of a specific model. In
fact, as it is apparent in its definition, Eq. (\ref{gamtil}), the
strength and the sign of ${\widetilde \Gamma}_\omega$ depend on the
balance between
$\Gamma_{\omega}^{e} (q,\omega) $ and $\gamma^2$. Moreover, as
we shall explicitly see in the model treated in the following
sections, $\Gamma_{\omega}^{e} (q,\omega) $ in turn results from a
cancellation between the strong bare repulsion and the strong
interband screening. How much is left from this cancellation
depends on the  specific model one is dealing with.

\section{The model}
The model we consider is represented by the following Hamiltonian
\begin{eqnarray}
H& = &\varepsilon_d^0 \sum_{i\sigma} d_{i\sigma}^\dagger
d_{i\sigma} +
\varepsilon_p^0 \sum_{i\sigma} (p_{i\sigma x}^\dagger p_{i\sigma x}
 + p_{i\sigma y}^\dagger p_{i\sigma y})
+ \sum_{<ij>\sigma}(t_{ij}
d_{i\sigma}^\dagger p_{j\sigma x}+x\rightarrow y+{\rm{h.c.}})
\nonumber\\
 & + &  \sum_{<ij>\sigma}(t_{ppij} p^{\dagger}_{i\sigma x}
 p_{j\sigma y}
+ {\rm{h.c.}}) + U_d\sum_{i}
 n_{di\uparrow} n_{di\downarrow}
+\omega_0\sum_iA^\dagger_iA_i \nonumber \\
& -& \sum_{i, \sigma} \left( A_i + A_i^\dagger \right)
\left[ G_d \left( n_{di}-\langle n_{di} \rangle \right)+
G_p \left( n_{pi}-\langle n_{pi} \rangle \right) \right].
\label{ham1}
\end{eqnarray}
where $\varepsilon_d^0$ and $\varepsilon_p^0$
are  the Cu and O energy levels  respectively,
$t_{ij} = \pm t_{pd}$ is the Cu-O hybridization,
$t_{ppij}=\pm t_{pp}$ is the nearest neighbor O-O  hybridization
(for the choice of the  orbital phases  and the
related choice of the sign of the hopping constants, see, e.g.,
Ref.\cite{GKM}). $U_{d}$ is the on site repulsion between holes on
copper sites. Starting from a Cu($3d^{10}$)-O($2p^6$) vacuum state,
holes on copper $d_{x^2-y^2}$ orbitals or on oxygen $p_{x}$ or $p_y$
orbitals at site $i$ are created by the
$d_i^\dagger$ and $p_i^\dagger$
operators respectively.
$n_{di}=\sum_\sigma d^{\dagger}_{i \sigma} d_{i \sigma}$
is the total density per cell of holes on copper, while
$n_{pi}=\sum_{\sigma, \alpha= x, y} p^{\dagger}_{i, \sigma ,\alpha}
p_{i, \sigma, \alpha}$ is the total density per cell of holes on
oxygen. The boson creation operator $A^\dagger_i$
creates a dispersionless phonon with frequency $\omega_0$ coupled to
the local density of copper and oxygen holes by the coupling
constants $G_d$ and $G_p$ respectively \cite{notag}.
These couplings can arise from
the dependence of the interionic Coulombic repulsion on the
relative position as well as on the hole occupation. An
estimate can be obtained when only the nearest neighbor Coulomb
interaction  $V(|r_i-r_j|)n_{di}n_{pj}$ is considered. Then,  a
first order expansion on the ion displacement and on the charge
fluctuations leads to an e-ph coupling of the form  appearing in
the  Hamiltonian (\ref{ham1}), with
\begin{equation}
G_d \propto \left. {\partial V \over \partial r_i} \right|_R
\langle n_{pj} \rangle\,;\,\,\,
G_p \propto \left. {\partial V \over \partial r_j} \right|_R
\langle n_{di} \rangle \label{gs}
\end{equation}
where $R$ is the equilibrium distance between nn ions.
Therefore, two different constants,  $G_d$ and $G_p$, have been
introduced owing to the different average occupations of the $d$
and $p$ orbitals  and to the difference in ionic masses entering the
standard normalization factors implicitly included in the
definition of the $G$'s. However, we do not want to stress
very much in the present context the relevance of the specific
form of the e-ph coupling appearing in Eq.(\ref{gs}) \cite{zeyher2}.
In fact we do not address the specific aspects (like,
e.g., symmetries
and strengths) of the phonons in the cuprates, our issue being
the understanding of the general properties of an Holstein
\cite{holstein} phonon in a strongly correlated system.

A similar three-band Hubbard model
was considered in Ref.\cite{kim}, where, however, an
intersite ``covalent" e-ph coupling was considered arising from
the ion-position dependence of the hopping integrals.

 Since our investigation concerns the interplay
between  strong interactions and the phonons we take the $U_d \to
\infty$ limit. In a standard way we handle the no-double-occupancy
constraint on copper sites by means of the slave-boson technique
\cite{barnes}-\cite{slabos}. Therefore, after performing the usual
substitution ${d}^{\dagger}_{i \sigma}\rightarrow d^{\dagger}_{i
\sigma}b_i, \,\,\, {d}_{i \sigma}\rightarrow b^{\dagger}_i d_{i
\sigma}$ the constraint becomes $\sum_\sigma d^{\dagger}_{i\sigma}
d_{i\sigma} +b^{\dagger}_ib_i = 1$. To equip the model with a formal
small expansion parameter, we introduce a standard large-N expansion
\cite{coleman}, where the spin index $\sigma$ runs from 1 to $N$.
The constraint is relaxed to assume the form
$\sum_\sigma d^{\dagger}_{i\sigma}
d_{i\sigma} +b^{\dagger}_ib_i = {N \over 2}$ and
the suitable rescaling of the hopping
$t_{pd} \rightarrow {t_{pd} / {\sqrt{N}} }$ must, in this
model, be joined by the similar rescaling of the hole-phonon
coupling $G \rightarrow {G/ {\sqrt{N}}}$ in order to
compensate for the presence of $N$ fermionic degrees of freedom.
 Once these transformations are
carried out the partition function of the final model can
 be written as a functional integral
\begin{eqnarray}
 Z & = & \int Dp^{\dagger}_{\alpha \sigma}Dp_{\alpha \sigma}
Dd^{\dagger}_{\sigma}Dd_{\sigma} Db^{\dagger}Db
D\lambda DA DA^\dagger e^{-\int_0^\beta Sd\tau}, \label{funcint}\\
  S & = & \sum_i \left[ \sum_{\sigma} d^{\dagger}_{i \sigma}
 {{\partial d_{i\sigma}} \over {\partial \tau}} +\sum_{\sigma
\alpha =x,y} p^{\dagger}_{i \sigma \alpha}
{{\partial p_{i\sigma \alpha}} \over {\partial \tau}}
 +b^{\dagger}_{i} {{\partial b_{i}} \over {\partial \tau}}
+A^{\dagger}_{i} {{\partial A_{i}} \over {\partial \tau}} \right]
+ \sum_i \left[ i\lambda_i \left( b^{\dagger}_i b_i
-{N\over 2}\right)  \right] +H,  \label{action}\\
 H & = & \sum_{i,\sigma} d^{\dagger}_{i \sigma} d_{i \sigma}
\left( \varepsilon^0_d +i\lambda_i\right)
 + \varepsilon^0_p \sum_{i,\sigma,\alpha=x,y}
p^{\dagger}_{i, \sigma ,\alpha} p_{i ,\sigma ,\alpha} \nonumber \\
   & - & {t_{pd} \over {\sqrt {N}} } \sum_{i, \sigma}
\left[ \left( p^{\dagger}_{i, \sigma ,x}-p^{\dagger}_{i ,\sigma, -x}
+p^{\dagger}_{i, \sigma, y}- p^{\dagger}_{i, \sigma ,-y}\right)
d_{i \sigma} b^{\dagger}_i + c.c.\right] \nonumber  \\
 & - & t_{pp} \sum_{i \sigma} \left[ p^{\dagger}_{i+x, \sigma, x}
\left( p_{i ,\sigma, -y} -p_{i, \sigma, y}+ p_{i+2x, \sigma, y}
-p_{i+2x, \sigma ,-y} \right) + c.c. \right] \nonumber \\
& -& {1 \over {\sqrt{N}}}\sum_{i, \sigma} \left( A_i + A_i^\dagger
\right)  \left[ G_d \left( n_{di}-\langle n_{di} \rangle \right)+
G_p \left( n_{pi}-\langle n_{pi} \rangle \right) \right]+
\omega_0\sum_iA^\dagger_iA_i.
\label{ham2}
\end{eqnarray}
where a local Lagrange multiplier field $\lambda$ has been
introduced to implement the local constraint forbidding the
double occupancy on copper.

 At the mean-field ($N=\infty$) level, the model
of Eqs.(\ref{funcint})-({\ref{ham2}) is
equivalent to the  standard, purely
electronic three-band Hubbard model without coupling to the
phonons, which has been  widely considered in the literature
\cite{KLR},\cite{GKM}. In fact, at mean field level no
role is played by the phonons because our electron-lattice coupling
depends on the difference between the local  and the average
density and this difference naturally vanishes in the mean-field
approximation \cite{notaeqil}.
 The model displays a T=0 Fermi-liquid
behaviour for any finite doping $\delta$, where $\delta$ is the
deviation from  half-filling, when one hole per cell is present in
the system. In the Fermi-liquid case the  mean-field value of the
slave-boson field $b_0$ multiplicatively
renormalizes the hopping thus enhancing the effective mass of the
quasiparticles ($b_0/\sqrt{N} \le \sqrt{1/2}$). Moreover, at this
level  the single particle self-energy does not introduce a finite
quasiparticle lifetime. Then, in this
model the single particle Green function of the physical  fermions
at $N=2$ has a quasiparticle pole with a finite residue given by the
square of the  mean-field value of the slave-boson field $b_0^2$.

On the other hand, at half-filling the system becomes an
insulator if the bare charge-transfer energy difference
$\varepsilon_p^0-\varepsilon_d^0-4t_{pp}$ is
larger than a critical value
ranging from $3.35t_{pd}$ when $t_{pp}=0$ \cite{KLR}
to a smaller value $\approx 2.5 t_{pd}$ when $t_{pp} =0.5
t_{pd}$ \cite{GKM}.
In the insulating phase $b_0$ vanishes leading to an infinite
quasiparticle mass ($m^*/m=\infty$)  and to a vanishing
quasiparticle spectral weight \cite{notamit}.

In order to get new physical effects from the presence of the
coupling with the phonons, one needs to consider the fluctuations
of the bosonic fields. Since only a particular combination
$a=(A^{\dag}+A)/(2\sqrt{N})$ of the phonon fields $A$ and
$A^{\dag}$ is coupled to the fermions, it is convenient to
use the field $a$ and to integrate out the orthogonal
combination ${\widetilde{a}}=(A-A^{\dag})/(2\sqrt{N})$.
Then the quadratic action for the boson field $a$ reads
\begin{equation}
H_{\rm {phon}}=N\sum_{n,i}
{{\omega_n^2+\omega_0^2}  \over {\omega_0}}a_i^\dagger
a_i\,\,, \label{aaction}
\end{equation}
where we have transformed the imaginary time into Matsubara
frequencies. Moreover, working in the radial gauge \cite{RN}, the
phase of the field  $b_i=\sqrt{N}r_i\exp(-i\phi) $ is gauged away
and only the modulus field $r_i$ is kept, while $\lambda_i$
acquires a time dependence $\lambda_i\to \lambda_i+\partial_\tau
\phi_i$.
Thus one can define a three-component field  $A^{\mu}=(\delta
r,\,\,\, \delta \lambda, \,\,\, a)$ where  the time- and
space-dependent components are  the fluctuating part of the boson
fields $r_i = r_0 \left( 1+\delta r_i   \right)$, $\lambda_i =
-i\lambda_0 + \delta \lambda_i$ and $a_i$. Writing the Hamiltonian of
coupled fermions and bosons as $H=H_{MF}+H_{\rm {bos}}+H_{\rm
{int}}$, where $H_{MF}$ is the mean-field Hamiltonian quadratic in
the fermionic fields, $H_{\rm{bos}}$ is the purely bosonic part, also
including the terms with the $a$, $r$ and $\lambda$ bosons appearing
in the action (\ref{action}) and in $H_{\rm {phon}}$,
Eq.(\ref{aaction}). $H_{\rm{int}}$  contains the fermion-boson
interaction terms. More explicitly, Fourier transforming to the
momentum space, the bosonic part reads  $$
H_{\rm {bos}}=N \sum_{q \mu \nu} A^{\mu}(q)B^{\mu \nu}(q)
A^{\nu}(-q)
$$
without explicitly indicating the frequency dependence for the
sake of simplicity. The form of Eqs.(\ref{action})-(\ref{aaction})
allows to determine the matrix $B^{\mu,\nu}$, whose elements are
all zero except for $B^{1,1}=r_0^2\lambda_0$,
$B^{1,2}=B^{2,1}=ir_0^2$, $B^{3,3}=\left( \omega_n^2+\omega_0^2
\right)/\omega_0 $.

To simplify the notation in $H_{MF}$ and in $H_{\rm {int}}$, we
introduce a three-component fermionic field ${\Psi}_{k
\sigma\alpha} \equiv \left( d_{k\sigma},
ip_{xk\sigma},ip_{yk\sigma} \right)$ describing in momentum space
the fermions in the orbital basis. Then $H_{MF}$ can compactly be
written as  $H_{MF}= \sum_{k \sigma \alpha \beta} H_{MF}^{\alpha
\beta}(k) {\Psi}^{\dagger}_{k \sigma \alpha} {\Psi}_{k \sigma\beta}$,
where $$
H_{MF}\left( k \right)=\left(
\begin{array}{c c c}
\varepsilon_d
& -2r_0t_{pd} \sin (k_x/2) & -2r_0t_{pd} \sin (k_y/2)  \\
-2r_0t_{pd} \sin (k_x/2)
&  \varepsilon_p          & -2t_{pp} \beta_k \\
-2r_0t_{pd} \sin (k_y/2) &    -2t_{pp} \beta_k     &   \varepsilon_p
\end{array} \right)
$$
with $\beta_k \equiv 2\sin(k_x/2)\sin(k_y/2)$
and $\varepsilon_d=\varepsilon_d^0+\lambda_0$ being the
mean-field-renormalized
energy level of the copper $d$ orbitals. The
above matrix can be put in diagonal form by a unitary
transformation to the quasiparticle basis $\tilde{\Psi}_{k \sigma
\alpha} = \sum_\beta U_{\alpha \beta}(k) {\Psi}_{k \sigma \beta}$
leading to $H_{MF}=  \sum_{k \sigma \alpha } E_{\alpha}(k)
\tilde{\Psi}^{\dagger}_{k \sigma \alpha} \tilde{\Psi}_{k
\sigma\alpha}$, where $E_{1,2,3}(k)$ are the mean-field-renormalized
 quasiparticle bands (we choose the band 1 as the
lowest one, so that the Fermi level lays in this band for filling
$\delta < 1$). This formalism allows to write the boson-fermion
interaction as \begin{eqnarray}
H_{\rm {int}} & = &
\sum_{k,q,\sigma}\Psi^{\dagger}_{k+{q\over 2}\sigma}
\Lambda^\mu\left(k,q\right)\Psi_{k-{q\over 2}
\sigma}A^\mu\left( q \right) \nonumber \\
 & = &
\sum_{k,q,\sigma}\tilde{\Psi}^{\dagger}_{k+{q\over 2}\sigma}
\tilde{\Lambda}^\mu\left(k,q\right)
\tilde{\Psi}_{k-{q\over 2}\sigma}A^\mu\left( q \right).
\end{eqnarray}
The fermion-component index has been dropped and the
$(3 \times 3)$ boson-fermion interaction vertices $\Lambda^\mu$
in the orbital operator basis can be obtained from Eq.(\ref{ham2})
$$
\Lambda^1= -2r_0t_{pd} \left(
\begin{array}{c c c}
0   &  \sin \left( {{k_x-q_x/2} \over 2} \right)  &
\sin \left( {{k_y-q_y/2} \over 2} \right)  \\
\sin \left( {{k_x+q_x/2} \over 2}  \right) & 0  & 0 \\
\sin \left( {{k_y+q_y/2} \over 2} \right)  & 0  & 0
\end{array} \right),
$$
\begin{equation}
\Lambda^2=  \left(
\begin{array}{c c c}
 i             &         0           &       0       \\
 0             &         0           &       0       \\
 0             &         0           &       0
\end{array} \right),\,\,\,\,\,
\Lambda^3=  \left(
\begin{array}{c c c}
 -2G_d               &             0           &       0       \\
 0                   & -2G_p\cos {q_x \over 2} &       0       \\
 0                   &         0               & -2G_p\cos {q_y
\over 2} \end{array} \right)\,\, ,
\end{equation}
\noindent
while the quasiparticle vertices
$\tilde{\Lambda}^\mu_{\alpha \beta}\left( k,q \right)$
are defined as
\begin{equation}
\tilde{\Lambda}^\mu\left( k,q \right)=
U\left( k+{q \over 2} \right) \Lambda^\mu\left( k,q \right)
U^{\dagger}\left( k-{q\over 2} \right) \label{Lamtil}.
\end{equation}
Introducing the fermionic bubbles coupled to the various bosons
\begin{equation}
\Pi^{\mu \nu}(q,\omega_m)=
\sum_{k,\alpha,\beta} {
{f\left( E_\alpha(k+{q\over 2}) \right)
-f\left( E_\beta(k-{q\over 2}) \right)} \over
{ E_\alpha(k+{q\over 2}) - E_\beta(k-{q\over 2}) -i \omega_m}}
\tilde{\Lambda}^\mu_{\alpha \beta}\left( k,q \right)
\tilde{\Lambda}^\nu_{\beta \alpha }\left( k,-q \right). \label{pi}
\end{equation}
one can define the boson propagator
\begin{eqnarray}
D^{\mu \nu}(q,\omega_m) & = &
<A^\mu(q,\omega_m)A^\nu(-q,-\omega_m)> \nonumber \\
 & = & N^{-1}(2B+
\Pi(q,\omega_m))^{-1}_{\mu \nu}
\label{bosprop}
\end{eqnarray}
The factor 2 multiplying the boson matrix B arises from the fact that
the bosonic fields in the presently used radial gauge are real.

The set of formal tools is then completed by the introduction of the
effective scattering amplitude between the quasiparticles
in the lowest band
\begin{equation}
\Gamma (k,k';q,\omega)= -\sum_{\mu \nu}
\tilde{\Lambda}_{11}^{\mu}  \left(k', -q
\right) D^{\mu \nu} \left( q, \omega \right)
\tilde{\Lambda}_{11}^{\nu} \left(k, q \right). \label{gamma}
\end{equation}
Then the scattering amplitude in the Cooper channel is given by
\begin{equation}
\Gamma^C(k,k';\omega) = -\sum_{\mu \nu}
\tilde{\Lambda}_{11}^{\mu}\left({k+k' \over 2} ,-k+k'\right)
D^{\mu \nu}(q=k-k',\omega)
\tilde{\Lambda}_{11}^{\nu}\left(-{k+k' \over 2},k-k'\right)
\label{cooper}
\end{equation}
It should be noted that the boson propagators are of order 1/N
while the occurrence of a bare fermionic bubble leads to a
spin summation and is therefore associated with a factor  N.
Thus, in this 1/N approach, the quasiparticle scattering
amplitudes are residual interactions of order 1/N. The matrix
form of the static  density-density correlation function at the
leading order is
\begin{eqnarray} P_{\alpha \beta}(q,\omega=0) & = & {1\over N}
\sum_{\sigma \sigma'}<n_{\alpha\sigma}(q) n_{\beta\sigma'}(-q)>
\nonumber \\
& = & P^{0}_{\alpha \beta}(q,\omega=0)+
N\sum_{\mu \nu} \chi^{0}_{\alpha \mu}(q,\omega=0) \  D^{\mu
\nu}(q,\omega=0) \
\chi^{0}_{\nu \beta}(q,\omega=0) \,\,\,\,\,\,\,\,\,
\label{pab}
\end{eqnarray}
where
\begin{equation}
P^{0}_{\alpha \beta}(q,\omega) = {1\over N}
\sum_{\sigma \sigma'}<n_{\alpha\sigma}(q) n_{\beta\sigma'}(-q)>_{0}
\label{p0ab}
\end{equation}
is the orbital bare density-density correlation function, and
 \begin{equation}
\chi^{0}_{\alpha \mu}(q,\omega) = {1\over N}
\sum_{\sigma \sigma'} <n_{\alpha\sigma }(q) {\sum_{k;\gamma,\delta}
\Psi^{\dagger}_{k\sigma'\gamma}\Lambda^{\mu}_{\gamma\delta}(k,q)
\Psi_{k+q\sigma' \delta} }>_{0}
\label{p1ab}
\end{equation}
where $\alpha=d,p_x,p_y$, and $\mu = 1,2$ and $3$.

\section{The dynamical analysis}
We now analyze the interplay between
lattice and holes in the strongly correlated system
represented by the model of Eqs.(\ref{funcint})-(\ref{ham2})
focusing on the possible occurrence of instabilities.

Two mechanisms for driving a system to an instability can in
principle be devised. A first one requires the complete softening
of a collective mode leading to a ground state with different
symmetries. For instance, in our specific model, this
mechanism could imply the softening of the phonon leading
to a structural transition.  To investigate this mechanism
a dynamical analysis of the collective modes of the system
will be carried out in Subsection A.

However, as we generally found out in Section II, an instability
can also  occur when the Landau criterion for stability $F_0^s > -1$
is violated (and usually a PS will then take place).
In this case, the instability shows up as an
overdamping of the zero-sound mode, the other collective modes
remaining massive, and is a result of
the delicate unbalance between the repulsive and the attractive
forces present in the system. Subsections  B and C are devoted
to this particular aspect, focusing on the   role of the various
screening processes in determining the instability.

The result of the analyses performed in Subsections A, B and C
is that the model of Eqs.(\ref{funcint})-(\ref{ham2}) undergoes
an instability driven by the second of the above mechanisms, and,
in general, no phonon softening is detected. A similar phenomenon
\cite{GRCDK1}-\cite{RCGBK} occurs in the three-band Hubbard model
in the presence of a nearest neighbor Coulombic repulsion. In this
latter case it is the attraction due to copper-oxygen CT fluctuations,
which determines a PS instability without being accompanied
by any collective-mode softening.

\subsection{The collective modes}
The collective modes appear as poles of the  density
susceptibilities.
However an important remark is that the
poles of the density susceptibilities $\chi$ coincide with the
poles of the boson propagator  $D^{\mu \nu} (q, \omega)$, which
therefore contain all of the relevant informations on the
collective modes. In particular, to find the dispersion
of these modes $\omega=\omega(q)$, one has to solve the equation
\begin{equation}
\det \left(2B+\Pi \right) =0. \label{det0}
\end{equation}
Quite generally the charge-density fluctuations in the three-band
model  can be decomposed into the fluctuations of a non-conserved
field, $n_d-n_p$, coupled to a fluctuating conserved field,
$n_d+n_p$. The
mode describing the propagating total density fluctuations
is the zero-sound mode. This mode is massless, since  in the model
of Eqs.(\ref{action})-(\ref{aaction}),  no long-range interaction
among the holes is included, which would otherwise transform this
mode into a massive plasmon. In the presence of an overall attraction
($0\ge F_0^s >-1$), the zero-sound mode enters into the particle-hole
continuum and becomes damped.
On the other hand the p-d CT
fluctuations, being described by a non-conserved field, will
contribute to the CT "optical-like" mode.  Of course these
fluctuations are dynamically coupled to the phonon mode.

In order to simplify our analysis
we first consider the $q=0$ limit because in this limit
all intraband fermionic bubbles vanish  thereby leaving the boson
propagator in a much more treatable form. Physically this is due
to the fact that  in the $q=0$ limit the total density field
$\rho (q) =n_d(q)+n_p(q)= (1/\sqrt{N_s})
\sum_k \tilde{\Psi}^{\dagger}_{k+{q\over 2}\sigma
}\tilde{\Lambda}^+ (k,q) \tilde{\Psi}_{k-{q\over 2} \sigma }$
with
\begin{equation}
\Lambda^+=  \left(
\begin{array}{c c c}
 1                   &         0           &       0       \\
 0                   & \cos {q_x \over 2}  &       0       \\
 0                   &         0           & \cos {q_y \over 2}
\end{array} \right) \label{vertices}
\end{equation}
\noindent
decouples from the dynamics because of particle conservation.
To allow for a simpler analytic treatment
we neglect in this section the direct  oxygen-oxygen overlap
$(t_{pp}=0)$. Thus the boson-quasiparticle vertices simplify to the
form \begin{eqnarray}
\tilde{\Lambda}^\mu_{12}  (k,q=0) & = & {{2r_0t_{pd}\gamma_k} \over
{R_k}} \left( \begin{array}{c}
\Delta  \\
i       \\
2G
\end{array} \right) \nonumber \\
\tilde{\Lambda}^\mu_{13} (k,q=0) & = & {{2r_0t_{pd}\gamma_k}
\over {R_k}} \left( \begin{array}{c}
0  \\
0  \\
2G
\end{array} \right)\, ,
\end{eqnarray}
where $\Delta\equiv\varepsilon_p^0-\varepsilon_d$ is the
mean-field-renormalized atomic
level difference and $G\equiv G_p-G_d$.
Then the following expression for the polarization bubble results
$$
\Pi^{\mu \nu} (q=0, \omega) = -2 r_0^2
\left(
\begin{array}{c c c }
\Delta^2  & i \Delta & 2 \Delta G  \\
i \Delta  & -1       & 2i       G  \\
2 \Delta G& 2i  G    & 4       G^2
\end{array} \right) I \left( \omega \right)
$$
where
$$
I\left( \omega \right) = {1 \over {N_s}} \sum_k
{{4t_{pd}^2\gamma_k^2} \over {R_k \left( R_k^2-\omega^2\right)}}
$$
Using Eq.(\ref{bosprop}) one  obtains the following
expression for  $D^{-1}$
$$
D^{-1}(q=0, \omega) = 2Nr_0^2
\left(
\begin{array}{c c c }
\lambda_0-\Delta^2I  & i(1-\Delta I) & -2 \Delta IG \\
i (1- \Delta I)      & I             & -2i IG       \\
-2 \Delta IG         & -2i IG        & -4IG^2+
{{\omega_0^2-\omega^2}\over {r_0^2\omega_0}}
 \end{array} \right)
$$
and the determinant can be evaluated as
\begin{equation}
\det {1\over N} D^{-1}\left( 0,\omega \right) =
\left(2r_0^2\right)^3
\left[
{{\omega_0^2-\omega^2}\over{\omega_0}}
\left(1- I\left(2\Delta-\lambda_0 \right) \right)
-4G^2I
\right].             \label{detbosprop}
\end{equation}
The analytical treatment can then proceed further introducing
the small $r_0^2$ approximation. This is justified at
low doping deep inside the region of
parameters where the system is insulating at half-filling.
We recall that the model we are considering at leading order in $1/N$
has a vanishing $r_0^2$ approaching the insulating regime.
Then it is convenient to rewrite $I(\omega)$ as
\begin{equation}
I(\omega ) = {\lambda_0 \over {\Delta^2-\omega^2}}
\left[ 1- \alpha (\omega )r_0^2 \right]
\end{equation}
with
$$
\alpha (\omega )\equiv {1 \over {\lambda_0 N_s}} \sum_k
{ {64t_{pd}^4\gamma_k^4} \over {R_k \left( R_k^2 -\omega^2 \right)} }
$$
so that the determinant assumes the form
\begin{eqnarray}
\det {1\over N} D^{-1}\left( 0,\omega \right) & = &
 {{8r_0^4} \over {\omega_0 \left(\Delta^2-\omega^2\right)}}
\left[
\left(\omega_0^2-\omega^2\right)
\left[ \left(\Delta-\lambda_0\right)^2 -\omega^2 \right]
\right. \nonumber \\
                                              & + &
\left. \lambda_0r_0^2 \left[  -4G^2\omega_0 +
\left(\omega_0^2-\omega^2\right) \left(2\Delta-\lambda_0\right)
\alpha(\omega)
\right]
\right].
\end{eqnarray}
By expanding $\alpha$ at zeroth order in $r_0^2$
\begin{equation}
\alpha (\omega) \approx {{16t_{pd}^2 \gamma_4} \over {\gamma_2} }
{ 1 \over { \Delta^2-\omega^2}} \label{alpha1}
\end{equation}
with $\gamma_{2} \equiv (1/N_s) \sum_k \gamma_k^{2 }=1/2+2/\pi^2
\simeq 0.702 $
and $\gamma_{4} \equiv (1/N_s) \sum_k \gamma_k^{4}=5/8+4/\pi^2
\simeq 1.03$
one obtains
the expression of the determinant at first order in $r_0^2$.
It is immediate to recognize that, taking the $r_0\to 0$ limit,
one obtains two collective modes
\begin{eqnarray}
\omega^2& = &
\left(\Delta-\lambda_0\right)^2 \equiv \omega_{exc}^2 \\
\omega^2 & = & \omega_0^2
\end{eqnarray}
corresponding to the excitonic charge-transfer mode
($n_p-n_d$ fluctuations) and to the bare phonon mode.
At this leading level in the 1/N expansion the two modes
do not mix. On the other hand the equation $\det D^{-1}=0$
can be solved giving
\begin{eqnarray}
\omega^2  & = &
{1\over 2}
\left[
\omega_0^2+\omega_{exc}^2 +
r_0^2 C\left({\overline{\omega}}\right)\lambda_0 \right. \nonumber \\
          & \pm &
\left.
\left[
\left( \omega_0^2-\omega_{exc}^2 \right)^2+
\lambda_0 r_0^2
\left[
2C\left({\overline{\omega}}\right)
\left(\omega_{exc}^2 -\omega_0^2 \right) +16G^2 \omega_0
\right]
\right]^{1 \over 2} \right] \nonumber \\
& \simeq &
{1\over 2} \left[
\left(\omega_0^2+\omega_{exc}^2\right)\pm
\vert\omega_0^2-\omega_{exc}^2 \vert \right. \nonumber \\
& +  &
\left. r_0^2
\left[
C\left({\overline{\omega}}\right)\lambda_0
\pm {\lambda_0 \over{\vert \omega_0^2-\omega_{exc}^2 \vert}}
\left[C\left({\overline{\omega}}\right)
\left(\omega_{exc}^2-\omega_0^2\right)
+8G^2\omega_0
\right]
\right]
\right]
\label{modes}
\end{eqnarray}
where the short notation $C\left({\overline{\omega}}\right)
=\left(2\Delta-\lambda_0\right)
\alpha\left(\omega={\overline{\omega}}\right)$  was introduced
and the result was expanded at first order in $r_0^2$.
${\overline{\omega}}$ assumes the values $\omega_{exc}$
or $\omega_0$ corresponding to the zeroth order term in the
expansion of $\omega$ in $r_0^2$.

Two cases can be distinguished in Eq.(\ref{modes}):
$\omega_{exc}>\omega_0$, the more physical one, and
$\omega_{exc}<\omega_0$ which  can only occur close to the
metal-charge-transfer-insulator transition, where the exciton
mode completely softens \cite{RCGBK},\cite{CKRGWR}.
In the former case one has
\begin{eqnarray}
\omega^2 & =&  \omega_{exc}^2+r_0^2\lambda_0 \left(
C\left(\omega_{exc}\right) + {{4G^2\omega_0}\over
{\omega_{exc}^2-\omega_0^2}} \right) \\
\omega^2 & = &  \omega_0^2-r_0^2\lambda_0 {{4G^2\omega_0}\over
{\omega_{exc}^2-\omega_0^2}}
\end{eqnarray}
whereas in the latter case the modes are
\begin{eqnarray}
\omega^2 & = & \omega_{exc}^2+r_0^2\lambda_0 \left(
C\left(\omega_{exc}\right) - {{4G^2\omega_0}\over
{\omega_0^2-\omega_{exc}^2}} \right) \\
\omega^2 & = & \omega_0^2+r_0^2\lambda_0 {{4G^2\omega_0}\over
{\omega_0^2-\omega_{exc}^2}}.
\end{eqnarray}
It should be noted that in both cases a "level repulsion"
occurs so that the higher energy mode is pushed at even higher
energy, while the lower energy one is made softer by the
reciprocal interaction. However, it is important to stress the
fact that this $q=0$ analysis does not reveal any complete
softening of the collective modes in the limit of vanishing
$r_0^2$, where the energies of the modes reduce to their
``bare'' values $\omega_{exc}$ and $\omega_0$.
 This indicates that any occurrence of $q=0$ instabilities
in the small $r_0$ region of the system is not due to a softening
of the modes. Of course the above
analysis was confined to the $q=0$ case and, therefore, it does not
allow to draw any conclusion on the occurrence of a collective mode
softening  at a finite wavelength or at large $r_0$.
The investigation of this possibility
requires a numerical analysis which was indeed performed.
The result is that the static susceptibilities diverge
first at $q=0$ and the softening of the modes
at finite $q$ only takes place
inside an unstable region, characterized by $F_0^s <-1$.
This instability  occurs  as a consequence
of the repulsion vs attraction unbalance generally presented in
Section II, which we now elucidate for the specific three-band
model under  consideration.

\subsection{The Landau Fermi-liquid analysis}
In order to investigate the instability conditions of the model
we now turn to a Fermi-liquid analysis transposing our $1/N$
calculation to the standard formalism presented in Section II
and identifying the Landau amplitudes at leading order in $1/N$.
In particular, starting from the definition
(\ref{gamma}) of the singlet effective scattering
amplitude one can introduce the standard Landau amplitudes
$$
\Gamma_\omega (k,k')= -\lim_{\omega \rightarrow 0}\lim_{q
\rightarrow 0} \sum_{\mu,\nu}\tilde{\Lambda}_{11}^{\mu}
\left(k', -q \right) D^{\mu \nu} \left( q, \omega \right)
\tilde{\Lambda}_{11}^{\nu} \left(k, q \right).
$$
To carry out an analytical treatment we again assume $t_{pp}=0$.
Then, since the model only includes a direct copper-oxygen transfer
integral, $\Gamma_\omega (k,k')$ depends on $k$ and $k'$
only via $\gamma_k$ and $\gamma_{k'}$. Therefore, by taking the
quasiparticles at the Fermi surface (where $\gamma_k$=constant=
$\gamma_F$),
only the "zeroth" harmonic would be non zero and given by
\begin{equation}
\Gamma_\omega = -\lim_{\omega \rightarrow 0}\lim_{q \rightarrow 0}
\sum_{\mu,\nu}\tilde{\Lambda}_{11}^{\mu} \left(k_F , -q \right)
D^{\mu \nu} \left( q, \omega \right)
\tilde{\Lambda}_{11}^{\nu} \left(k_F, q \right).
\end{equation}
To explicitly get the expression of
$\Gamma_\omega$ we first calculate
$$
\Gamma (\omega)=- {1 \over {2Nr_0^2}}
\left(
\begin{array} {c}
-{ {R_{k_F}^2-\Delta^2} \over {2R_{k_F}} } \\
i{ {R_{k_F}+\Delta    } \over {2R_{k_F}} } \\
-2\left(G_du^2_{k_F}+G_pv^2_{k_F}\right)
\end{array} \right)
$$
$$
\times \left(
\begin{array}{c c c}
a                            & -i-ia(\Delta -\lambda_0)
& 2r_0a h(\omega)            \\

-i-ia(\Delta -\lambda_0)     & \lambda_0-(\Delta -\lambda_0)^2a
&-2ir_0^2(\Delta-\lambda_0)ah(\omega) \\
2r_0a h(\omega)          &-2ir_0^2(\Delta-\lambda_0)ah(\omega)
&r_0^2\omega_0\left(1+4r_0^2ah(\omega)\right)/
\left(\omega_0^2-\omega^2\right) \end{array} \right) \nonumber
$$
\begin{equation}
\times \left(
\begin{array} {c}
-{ {R_{k_F}^2-\Delta^2} \over {2R_{k_F}} } \\
i{ {R_{k_F}+\Delta    } \over {2R_{k_F}} } \\
-2\left(G_du^2_{k_F}+G_pv^2_{k_F}\right)
\end{array} \right) \label{gam0}
\end{equation}
with  $a\equiv a(\omega) \equiv I(\omega )
\left[1-\left(2\Delta-\lambda_0 +8r_0^2V \right)I(\omega )
\right]^{-1}$ and $h(\omega)=\omega_0G^2/(\omega_0^2-\omega^2)$
This gives
\begin{equation}
\Gamma (\omega)=\Gamma_0+a(\omega)\Gamma_1(\omega),
\end{equation}
with
\begin{eqnarray}
N\Gamma_0 & = & {1 \over 8r_0^2}\left(1+ {\Delta \over R_{k_F}}
\right)^2 \left(2R_{k_F}-2\Delta+\lambda_0 \right)- {2 {\omega_0}
\over
{\omega_0}^2-\omega^2} \left(G_du^2_{k_F}+G_pv^2_{k_F}\right)^2
\label{g0}\\
N\Gamma_1 & = & -{1 \over 2r_0^2}
\left[  {1 \over 2} \left(1+ {\Delta \over R_{k_F}} \right)
\left(R_{k_F}-2\Delta+\lambda_0 \right) \right. \nonumber \\
& + & \left. 4r_0^2{2 {\omega_0} \over {\omega_0}^2-\omega^2}
\left(G_d-G_p\right) \left(G_du^2_{k_F}+G_pv^2_{k_F}\right)
\right]^2. \label{g1}
\end{eqnarray}
A simple expression for  $\Gamma_{\omega}$ can be obtained taking the
$\omega \to 0$ limit of the above formulas in the small $r_0$
limit. It is worth noting that, whereas $a(\omega)$ is always finite,
 both $\Gamma_0$ and $\Gamma_1$ grow very large when $r_0\to 0$.
This is reminiscent of the infinite U we started with.
However a straightforward expansion of Eqs.(\ref{g0}) and
(\ref{g1}) shows that the leading terms of order $1/r_0^2$
(which characterize the model even in the absence of the
e-ph interaction)
cancel and only finite contributions are left.
Therefore in this model a finite effective scattering amplitude is
the result of a cancellation between a very large bare repulsion
and a very large attraction due to interband screening, to which
a finite phonon contribution is added.

The resulting dynamical scattering amplitude is, for $r_0 \to 0$
\begin{equation}
N\Gamma_{\omega}\approx {4\lambda_0 \over
{\left(\lambda_0-\Delta\right)^2}} \left[
2t_{pd}^2 \left({\gamma_4 \over \gamma_2}-1+ {\Delta^2 \over
4t_{pd}^2\gamma_2}\right)-
{\left(\lambda_0 G_p-\Delta G_d\right)^2 \over
2\lambda_0{\omega_0}}  \right]
\label{gamdyn}
\end{equation}
The first term represents the purely electronic
(interband screened) repulsive
contribution $\Gamma_\omega^{e}$ to $\Gamma_\omega$ \cite{RCGBK},
whereas the
second term is the attraction, which  arises from the e-ph
interaction. In particular it would coincide
with the second term in the r.h.s. of Eq.(\ref{gamtil}), if
the simple case $G_p=G_d=g$ is considered and the identification
$\gamma^2\equiv g^2/\omega$ is made \cite{notav}.
One can easily check that the e-ph vertex corrections
in Eq.(\ref{gamdyn})
(i.e. the factors $\Lambda_\alpha \Lambda_\beta$
with $\alpha,\beta =p,d$,
which multiply $G_\alpha G_\beta /\omega_0$)
can be written as
\begin{equation}
\Lambda^\omega_\alpha = {{\rm {d}} \phantom{\varepsilon^0_\alpha}
 \over {\rm {d}} \varepsilon^0_\alpha}
G^{-1}_{\rm {qp}} (k_F, \omega =0)\,\, ,
\end{equation}
where $G_{\rm {qp}}(k,\omega )=\left( \omega - E_k \right)^{-1}$
is the quasiparticle Green function and the derivatives are taken
at fixed doping $\delta$.
One can then discover that in a multiband model
the vertex corrections are still different from zero in the
dynamical limit (cf. Eq.(\ref{wi1}) with $z_w=1$ because we
are considering quasiparticles) and in fact
can act to enhance the vertices. For instance, at $\delta
=0^+$, we obtain $\Lambda^\omega_p= \lambda_0 /(\lambda_0-\Delta)$,
which reduces to unity only in the $\lambda_0 \to \infty$
limit, i.e. $\varepsilon_p^0-\varepsilon_d^0 \to \infty$.
To obtain the vertex correction
in the static limit we need to evaluate
${{\rm {d}} \phantom{\varepsilon^0_\alpha}
 \over {\rm {d}} \varepsilon^0_\alpha}
G^{-1}_{\rm {qp}} (k_F, \omega =0)$ at fixed chemical potential
\cite{notafix}. We obtain
\begin{equation}
\Lambda_\alpha^q={\Lambda_\alpha^\omega
\over 1+N\nu^*\Gamma^e_\omega}\,\, ,
\label{vst}
\end{equation}
where $\Gamma^e_\omega$ is the first term in the r.h.s. of
Eq.(\ref{gamdyn}). Eq.(\ref{vst}) shows the dramatic difference
between the dynamic and static limits. As discussed in Section II,
this is a generic feature of strongly correlated systems,
which one has to take into account since it can strongly affect
the relevance of the e-ph coupling \cite{notahub}.

\subsection{The phase diagram}
The previous analysis of the restricted model with $t_{pp}=0$
provides valuable informations on the possible occurrence of
instabilities of the Fermi liquid. In particular, having found
$\Gamma_\omega$, one can determine the Landau parameter $F_0^s\equiv
N\nu^*\Gamma_\omega$. Since the criterion for a finite
positive compressibility is
$F_0^s > -1$, and since $\nu^* \propto {1 \over r_0^2} \gg 1$ in the
small doping region, close to the
insulating region $(\varepsilon_p^0-\varepsilon_d^0>3.35 t_{pd})$
the stability criterion reduces to $\Gamma_\omega > 0$.
Taking into account the self-consistency condition
$\lambda_0 =4t_{pd}^2\gamma_2/\Delta$
one is lead to the following inequality
\begin{equation}
\left( \lambda_0 G_p-\Delta G_d \right)^2 < 4t_{pd}^2
\lambda_0{\omega_0} \left({\gamma_4 \over \gamma_2}-1+{\Delta \over
\lambda_0}\right) \label{stability}
\end{equation}
Even for a weak e-ph coupling, this condition can
easily be violated (see below).

A clearer understanding of the stability condition  can be obtained
by considering the phonon coupled separately with  the copper or
the oxygen local density. In the case of $G_d=0$ the condition
(\ref{stability}) becomes
\begin{equation}
{G_p^2 \over {\omega_0}} < {4t_{pd}^2 \over \lambda_0}
\left({\gamma_4 \over \gamma_2}-1+{\Delta \over \lambda_0}\right)
\label{gd0}
\end{equation}
whereas for $G_p=0$ one has
\begin{equation}
{G_d^2 \over {\omega_0}} <
{\lambda_0^2 \over \Delta^2} {4t_{pd}^2 \over
\lambda_0} \left({\gamma_4 \over \gamma_2}-1+{\Delta \over
\lambda_0}\right). \label{gp0}
\end{equation}
Then one can easily recognize that, in the positive doping case
with  $\lambda_0 > \Delta$, the coupling of the phonon with the
oxygen holes $G_p$ is more effective in driving the instability.
The role of $\lambda_0$ and $\Delta$ is interchanged at negative
doping so that $\lambda_0 < \Delta$  and
in this latter case the coupling
$G_d$ plays a major role in rendering
the Fermi liquid unstable. Notice, however, that the model itself
is not particle-hole symmetric so that the expressions (\ref{gd0})
and (\ref{gp0}) do not interchange
in going from negative to positive doping.

It is worth noting that, no matter how small the e-ph coupling is
or how large the phonon frequency is, the instability will
always take place by increasing the large bare charge-transfer
difference $(\varepsilon_p^0-\varepsilon_d^0)/t_{pd}$. This is
due to the fact that, for large
$(\varepsilon_p^0-\varepsilon_d^0)/t_{pd}$ one has
$\lambda_0 \gg \Delta$
for positive doping or $\lambda_0 \ll \Delta$ for
negative doping. Then it is possible to make the r.h.s. of
Eqs.(\ref{gd0}) or (\ref{gp0}) smaller than the l.h.s..

The estimates (\ref{gd0}),(\ref{gp0}) were derived for the
three-band Hubbard model with $t_{pp}=0$
close to the insulating regime, where $r_0 \approx 0$.
To investigate the finite-doping regime and the model with finite
oxygen-oxygen hopping we explicitly carried out a leading
order 1/N analysis of the static $q$-dependent
density-density response function
$\chi(q,\omega=0)\equiv <(n_{p}(q)+n_{d}(q))(n_{p}(-q)+n_{d}(-q))>$,
which gives the compressibility of the system once the $q\to 0$
limit is taken. This response function
is obtained as a linear combination of the orbital
density-density correlation functions in Eq.(\ref{pab}).
 A divergent $\chi(q,\omega=0)$ signals an instability of the Fermi
liquid occurring at a wavevector $q$.

We performed the calculations for many values of the parameters,
and specifically for various phonon frequencies $\omega_0$ and
various e-ph couplings $G_p$ and $G_d$, and we particularly
explored the range of parameters, which could be relevant for the
copper oxides. According to recent
estimates \cite{dewette},\cite{nozaki} we considered various values
of $\omega_0$ ranging in the interval between 0.01 and 0.08eV,
particularly focusing on the 0.02eV region, where the phonon
density of states is highest.
As far as the e-ph couplings are concerned,
these quantities are not directly accessible in experiments
and their theoretical calculation is not easy due to the strongly
interacting nature of the high temperature superconductors
and to the nesting occurring in their Fermi surface
\cite{krakauer}.
This is why  we explored many possible cases obtaining
qualitatively similar results. A reasonable choice is to use both
$G_p$ and $G_d$ different from zero, because a coupling of the
lattice with both oxygen and copper hole density is naturally
expected. The values of $G_d$ and $G_p$ were such that a reasonable
effective coupling $\lambda<1$ results far enough from the
instability region. Moreover the  choice $G_p > G_d$ appears
reasonable in the case of intermediate hole doping according to the
observations  reported in the paragraph after Eq.(\ref{ham1}): i)
$G_p$ is proportional to $\langle n_d \rangle$, while  $G_d$ is
proportional to $\langle n_p \rangle$, with $\langle n_d \rangle$
usually being larger than $\langle n_p \rangle$; ii) a factor
$1/\sqrt{M_O}$ is included in the definition of $G_p$, whereas an
analogous factor $1/\sqrt{M_{Cu}}$ enters in $G_d$ further
justifying the assumption $G_p>G_d$, since $M_{Cu}>M_O$. Both
theoretical calculations and experiments in superconducting
cuprates \cite{stanford} support the above qualitative arguments.
The fact that oxygen holes are more strongly coupled to the lattice
than holes on copper has relevant physical consequences in the
light of the  above observation that, for positive doping, $G_p$ is
more effective than $G_d$ in driving the instability.

 The phase diagrams in the
$(\varepsilon_p^0-\varepsilon_d^0)/t_{pd}$ vs $\delta$
plane resulting from the analysis of $\chi (q, \omega=0)$
are shown in Figs. 1 and 2 for typical parameter sets.
The instability line indicates where $\chi (q \to 0,\omega =0)$
diverges and it delimits a region of negative compressibility.
 The most impressive consequence of the coupling with the phonons is
that a large unstable region appears at large doping or at large
bare charge-transfer gap in both diagrams. It is worth noting again
that the instability is {\it {not\/}} related to any collective mode
softening. It can rather be interpreted as due to the attraction
arising from the e-ph interaction eventually resulting in the
overdamping of the zero-sound mode. Approaching the instability,
the zero-sound mode first enters the continuum and gets
Landau-damped. Eventually it becomes overdamped and the instability
takes place.

It is important to stress that the instability of
the phase diagrams in Figs. 1 and 2 is an electronic instability
even if it is driven by the coupling with the lattice. Moreover
we find that {\it {the instability of
$\chi (q,\omega=0)$ first occurs at $q \to 0$\/}} and is
therefore a signature of a long-wavelength, static, thermodynamical
PS region to be identified by a Maxwell construction. To make this
point more explicit, we report in  Fig.3 the Cooper-channel static
scattering amplitude between quasiparticles on the Fermi surface
$\Gamma^C (k_F,k_F', \omega=0)$
(see Eq.(\ref{cooper}) and cf. also the
$\omega =0$ limit of Eq.(\ref{totgam})) for a system with
finite $t_{pp}(=0.2t_{pd})$ and at various dopings. According
to the observation already reported at the beginning of
Subsection A, the scattering amplitude  carries information on the
instability because a divergency in this quantity  can only be due
to a divergency in  the boson propagator also entering the
expression for $\chi(q, \omega=0)$, with
$q=k_F-k_F'$. Thus a divergent scattering
amplitude directly signals  a divergent $\chi (q, \omega=0)$. As it
can be seen in Fig.3, the divergency can also occur at sizable
$q$'s inside the unstable region. This can have relevant physical
consequences, once a long-range Coulomb force is considered in the
model. This latter interaction would  stabilize the system
in the regions of the phase diagram where $\chi (q,\omega =0)$
shows divergencies at low-$q$. This would prevent
the occurrence of PS, but would leave
open the possibility of finite-$q$ instabilities: Most probably a
sizable unstable region would survive to the introduction of a
long-range Coulomb repulsion in the phase diagram parts where
$\chi(q,\omega=0)$ was diverging at sizable $q$'s, i.e.
 the divergency of the density-density response function
would first occur at a finite
$q$ leading to the formation of incommensurate
CDW. A similar phenomenon was first
suggested for a different model in Refs.\cite{GRCDK1}-\cite{RCGBK}
and seems to be confirmed by a static analysis of a pseudospin
model in  Ref.\cite{emerykivelson}. Of course the above argument
 misses the dynamical aspects of the problem, and should be
taken as purely indicative of a possible scenario.
One should also take into account that superconductivity
can take place near the instability line where $F_0^s=-1$
and act to stabilize the system (i.e. superconductivity
could partially or fully preempt the instability region).

Few comments on the difference between the two phase
diagrams of Figs.1 and 2 are in order. It is apparent that the
prominent effect of the direct oxygen-oxygen hopping is a
stabilization at low doping and large
$\varepsilon_p^0-\varepsilon_d^0$. In order to
clarify this  point we performed a calculation of the
compressibility both in the presence and in the absence of $t_{pp}$
in the limit of very large
$(\varepsilon_p^0-\varepsilon_d^0 )/t_{pd}$ and small
positive doping.
Due to the slave-boson-mean-field band renormalization
the lowest quasiparticle band closely approaches the bottom of the
intermediate band, mostly of oxygen character. This latter band
is a flat non-bonding oxygen band in the absence of $t_{pp}$
and acquires a dispersion $4t_{pp}$ in the presence of direct
oxygen-oxygen hopping if the mixing with the copper $d_{x^2-y^2}$
orbitals is neglected.
In the $(\varepsilon_p^0-\varepsilon_d^0 )/ t_{pd}\to \infty$ and
$\delta \to 0^+$ one has a vanishing distance between the
renormalized $d$ atomic level and the bottom of the pure-oxygen band
$\tilde{\Delta} \equiv \varepsilon_p^0-4t_{pp}-\varepsilon_d \to
0$. In this limit the calculation
greatly simplifies and, for $t_{pp} \ne 0$ the compressibility per
spin assumes the form
\begin{eqnarray}
\kappa & \propto & {1 \over t_{pp}}
\,\,\,\,\,\,\,\,\,\,\,\,\,\,\,d=2; \label{comp2} \\
\kappa & \propto & {1 \over t_{pp}}
\left({\tilde\Delta \over t_{pp}} \right)^{1\over 2}  \,\,\,\,\,\,
d=3. \label{comp3}
\end{eqnarray}
This result is  suggestive of a decoupling between the $p$ and the
$d$ levels occurring when
$(\varepsilon_p^0-\varepsilon_d^0 )/ t_{pd}\to \infty$:  the
compressibility for the  interacting system of $(1+0^+)$ holes in
the mixed $p$-$d$ quasiparticle band  beares resemblance with the
compressibility of a system of $\delta=0^+$ non-interacting holes
in the pure-oxygen band. In this latter system the compressibility
coincides with the density  of states, which is of order $1/t_{pp}$
and is independent on doping in two dimensions, whereas it vanishes
with doping in the $d=3$ case. This is precisely the result
reported in Eqs.(\ref{comp2}) and (\ref{comp3}) respectively, once
one remembers that, in the
$(\varepsilon_p^0-\varepsilon_d^0 )/ t_{pd}\to \infty$ and
$\delta \to 0^+$ limits, $\tilde{\Delta}$ is vanishingly small.

Eqs.(\ref{comp2}) and (\ref{comp3}) also express the fact that
in the large $(\varepsilon_p^0-\varepsilon_d^0 )/ t_{pd}$
and low-doping limit,
the compressibility of the system diverges when  $t_{pp}$ tends to
zero (even for the three-band Hubbard model in the absence of e-ph
coupling).
Since $\kappa^e=\nu^*/(1+N\nu^*\Gamma_\omega^e)$, this result
can be rephrased as $\Gamma_\omega^e \simeq t_{pp}$. Thus, when
$t_{pp}\to 0$ and the quasiparticle density of states $\nu^*$ is
large, even a very small attraction leads to a negative
$\Gamma_\omega$, immediately resulting in a violation of the Landau
stability condition $F_0^s=N\nu^*\Gamma_\omega<-1$.  The three-band
Hubbard model with $t_{pp}=0$ and
$\varepsilon_p^0-\varepsilon_d^0=\infty$ case is strictly
analogous to the $U=\infty$ single-band case \cite{zeyher}.
This clarifies why,
the region with infinite compressibility  at low doping
and large $\varepsilon_p^0-\varepsilon_d^0$ in the phase diagram
of the system of Fig.1 ($t_{pp}=0$),
is more extended than the one for the system of Fig.2.

\subsection{The Cooper instability}
Finally we investigated the possibility of Cooper pairing.
As already pointed out, superconductivity can appear as
a precursor of the PS instability. In particular a
simple argument can be put forward
\cite{GRCDK1},\cite{GRCDK2} suggesting that a region of large
 compressibility is a good candidate in order to find
superconductivity. In the general language of Landau Fermi-liquid
theory the compressibility can be written as $\kappa=\nu_0{ m^* \over
m} /\left( 1+F_0^s \right)$. Then, if the compressibility gets
large without being accompanied by a large mass enhancement,
(as it happens near the instability line), this
means that the denominator is becoming  small indicating a negative
$F_0^s$ (indeed $F_0^s=-1$ at the instability line). Although this
parameter is related to the quasiparticle scattering in the
particle-hole channel, it seems reasonable to expect attraction in
the particle-particle  channel as well. According with this
plausibility argument and according to the previous experience in
other strongly interacting models, we therefore investigated the
Fermi surface average of the particle-particle scattering amplitude
defined in Eq.(\ref{cooper})
$$
\lambda_l =- { {
\int \int dk dk' \delta \left( E(k)-\mu \right)
\delta \left( E(k')-\mu \right)
g_l(k) \Gamma^C \left( k,k';\omega=0 \right) g_l(k')} \over
{\int dk \delta \left( E(k)-\mu \right) g_l(k)^2} }.
$$
with $g_{s_1}(k)=\cos(k_x)+\cos(k_y)$, $g_{s_2}(k)=(\cos(k_x)-\cos(k_y))^2$,
$g_{d_1}(k)=\cos(k_x)-\cos(k_y)$ and $g_{d_2}(k)=\sin(k_x)\sin(k_y)$
projecting the interaction onto the s-wave and d-wave channels.
(Notice that $\lambda_l>0$ means attraction).
Whereas the couplings $\lambda_d$ are found to be generally
attractive near (and inside) the unstable region,
we find s-wave Cooper instabilities only very close to the
instability line.  The results are tabulated at various doping
concentrations for the case with $t_{pp}=0$
and  $(\varepsilon_p^0-\varepsilon_d^0)/t_{pd}=3.3t_{pd}$
in Table 1 and for the case with $t_{pp}=0.2t{pd}$
and  $(\varepsilon_p^0-\varepsilon_d^0)/t_{pd}=4.9t_{pd}$
in Table 2. In both cases the phonon frequency and the e-ph couplings
are $\omega_0=0.02t_{pd}$, and $G_p=0.15t_{pd}$, and $G_d=0.1t_{pd}$,
respectively. With the set of parameters related to Table 1
the critical doping for the occurrence of the instability
is $\delta_c=0.21$, whereas for the set of Table 2 the instability
is at $\delta_c=0.23$. Our analysis indicates the sure
existence of d-wave pairing in sizable  regions of model
(\ref{ham2}) in the limit of large-N near the instability, whereas
the occurrence of s-wave pairing takes place in a much
narrower region \cite{notals}. However, it should be
emphasized that the presence of a  s-wave static Cooper
instability  only in a narrow region, by no means excludes the
possibility of having s-wave superconductivity in a much larger
area of our phase diagram. Only an appropriate   Eliashberg
dynamical analysis can allow to draw a conclusion, specially in the
light of the  considerations on the strong
frequency dependence of the vertex corrections discussed in
Section II. Of course the same applies to the  attraction in the
d-wave channels, which could also be greatly favored by  dynamical
effects. The full Eliashberg dynamical analysis unfortunately
involves a difficult complete treatment of the momentum and
frequency dependence of the effective scattering amplitude,
which is beyond the scope of the present paper. Nevertheless
valuable informations can be gained from an analysis of
the e-ph vertex. The aim of this analysis is to provide useful
quantitative indications on the behaviour of the e-ph interaction
as a function of both momentum and imaginary frequency in various
parts of the phase diagram as a preliminary step towards the
implementation of the full Eliashberg treatment of the effective
phonon-mediated e-e scattering amplitude.
Also in this case we checked that our  results do
not qualitatively change by varying the model parameters.
In particular we choose the values of the various quantities in
order to allow for a possible  connection with the experimentally
known features of the copper oxides. We choose the value of the
bare atomic level difference
$\varepsilon_p^0-\varepsilon_d^0=4.9t_{pd}$ so that
a CT optical gap in the insulating  phase  of about $2t_{pd}$
would result. Assuming the usual value of $t_{pd}=$1.3eV
for the copper-oxygen hopping integral, this latter value
turns out to be 2.6eV, and is not much larger than the
experimentally known values of the CT gap in the insulating
phase of the superconducting copper oxides (e.g. in Ref.\cite{falk}
the reported value for ${\rm{La_{2}CuO_4}}$ is 2.3eV).
As values of the phonon frequency we take
0.02$t_{pd}\simeq$0.026 eV and
0.08$t_{pd}\simeq$0.104 eV.
The first value was chosen because in this frequency region
the experimental phononic density of states is large, whereas
the larger value was used to extend the region where dynamical
effects can take place before the phonon frequency cuts off
the effective interaction.
The value of the e-ph couplings
was chosen accordingly in order to obtain reasonable values of
$\lambda$ \cite{krakauer}.

We first report in Fig.5 the interaction (slave-boson)
renormalized  vertex between phonons and quasiparticles
{\it {without phonon processes}}
included in it (see Fig.4) as a function of the Matsubara
frequency. Two of these  quantities can be joined to  a bare phonon
propagator to give the one-phonon effective scattering amplitude
specializing to the three-band Hubbard model the general expression
(\ref{phoneffint}).

On the other hand, joining two of the above vertices  with a fully
renormalized phonon propagator leads to the full phonon-mediated
scattering amplitude in the Cooper channel.
This quantity was represented by the second
term of Eq.(\ref{totgam}) in Section II and is reported in Fig.6
for the three-band Hubbard model as a function of the Matsubara
frequency. The momenta of the external fermions correspond
to the Cooper channel with $k$ and $k'$ on the Fermi surface.

It is clear that the results reported in Figs.5 and 6
quantitatively confirm the general qualitative analysis  of
Section II. Fig.5 displays the strong dependence of
the one-phonon vertex from the momentum vs frequency ratio.
A typical bandwidth of about 0.2$t_{pd}$ was chosen,
giving a Fermi surface density of states $\approx 13/t_{pd}$.
We only report the results corresponding to a doping $\delta=0.225$
since  no significant dependence of this vertex from the
doping is detected, so that the closeness to the instability is
immaterial in this quantity. On the other hand, it is apparent that
a rapid increase occurs in the phonon-quasiparticle vertex, when
the frequency becomes larger than $v_F q$. This is most evident  at
low momenta, where the phonon-quasiparticle vertex $G_{\rm {eff}}$
increases fast by more than one order of magnitude, from the small
(see inset in Fig.5) value at zero
frequency up to a sizable value at large-frequency. This
latter is  much larger because at frequency larger than the
bandwidth for quasiparticle-quasihole excitations ($v_Fq$ at
small $q$ in an isotropic system)
the intraband electronic screening is ineffective (cf.
Eq.(\ref{wi1})) \cite{notakim}.

Analyzing in detail the contributions coming from the
vertex corrections, we discovered that in this ``high"-energy
region the {\it {interband \/}} screening acts
to enhance the vertex at small $q$'s by
a factor of about 1.5 with respect to the bare vertex
($G_0\simeq 0.2$). The bare value is only recovered
at very high energy (much larger than the interband
CT energy $\Delta$), and/or large $q$.

More dramatic are the momentum and
doping dependences of the attractive scattering
amplitude in Fig.6. In the stable region ($\delta=0.2$), a maximum
in the attraction is observed at finite frequency and low momentum.
This maximum is a result of the balance between the
energy dependence of the vertex correction and of the
boson propagator (see the inset). The interaction at low momenta
is by far larger than the attraction at large momenta. The  same
behavior as a function of momenta holds near the unstable region of
doping, ($\delta=0.225$), but at low energy and small momentum
the attraction is not suppressed by vertex corrections,
because in the vicinity to the instability line the attraction
reaches its maximum at $\omega=0$.

Finally Figs.7 and 8  report the total effective scattering
amplitude in the Cooper channel $\Gamma^C(k_F,k_F',\omega_n)$
(continuous lines) resulting from the sum of the attractive
amplitude (dotted lines)  of Fig.6 and the repulsion arising from
purely electronic scattering processes (dashed lines) as a
function  of Matsubara frequencies at small (a) and large $q$ (b).

Being the calculations performed close to the instability, we see
that the e-ph interaction gives a large contribution to the total
interaction at small $q$, leading to an overall attraction at low
frequencies. On the contrary, at large $q$ the full interaction is
only governed by the pure e-e repulsion and no phonon
energy scale is detectable in its frequency dependence.
Fig.8 reports the same quantities of Fig.7, but for
a larger phonon frequency (consequently $G_p$
has been increased, $G_p=0.3t_{pd}$, to drive the system unstable
at reasonable doping $\delta_c=0.185$).  Qualitatively the results
of Fig.8
show the same behaviour depicted in Fig.7, but it is evident
that the phonon-mediated  attraction extends over a
larger frequency range. We also found that in this
latter case the attraction persists up to larger
momenta. At large $q$ (Fig.8b) the results
for $\Gamma^C(k_F,k_F,\omega_n)$ are still very similar to those
of Fig.7b.

Finally Fig.9 reports $\Gamma^C(k_F,k_F',\omega_n)$
as well as its attractive and purely electronic repulsive components
at $\delta=0.2$. Due to the depressing effect
of the vertex  at low frequencies, the phonon contributes
very little to $\Gamma^C(k_F,k_F',\omega_n)$ even
at small $q$. Nevertheless, as it is also apparent from Fig.6,
at frequency of the order of $\omega_0$ the attractive part
contributes sizably to the total interaction. This clearly
indicates that
at small q, contrary to the static attraction, which is strongly
suppressed by the vertex corrections,
the finite frequency attraction persists in regions
of doping that are far from the instability line.

The analysis summarized in Figs.5-9 is of obvious pertinence
in a complete Eliashberg treatment of the superconductivity
problem. In particular it is evident that the huge enhancement
of the attractive part of the scattering amplitude
near the instability line can be
responsible for large critical temperatures despite the small
e-ph coupling.
The closeness to a PS instability appears therefore as
a favorable condition in order to obtain high temperature
superconductivity from a phonon-mediated attraction
similarly to what suggested in the context of purely
electronic pairing mechanisms.

\section{Conclusions}
In the present work we investigated the e-ph
interaction in the presence of a strong local
repulsion within a general Landau
Fermi-liquid framework. Using standard Ward
identities, we pointed out the strong dependence of the e-ph vertex
on the momentum vs frequency ratio in a strongly interacting system
displaying  a large effective-mass enhancement, but not too large
compressibility. In particular we
showed that the dimensionless attractive quasiparticle scattering
mediated by a single phonon exchange
coupled to the electronic density is strongly suppressed by
vertex corrections due to the e-e repulsion when $v_Fq > \omega$. On
the contrary the dimensionless effective one-phonon-mediated
attraction is strongly enhanced by the effective mass increase when
$v_Fq < \omega$. These results stay valid within a random-phase
resummation of phonon exchanges, provided the system is far from an
instability. This strong momentum vs frequency dependence of the
e-ph  coupling shows up in different ways in different physical
quantities. In particular it was found in a three-band Hubbard
model  with intersite "covalent" e-ph coupling \cite{kim}, that
phonons contribute little to the quasiparticle scattering as far as
transport properties are concerned.
We expect a similar depression for the e-ph coupling considered
in this paper \cite{notaopt}. This is so because transport
properties involve low-frequency-high-momentum processes, where the
large-momentum limit of the e-ph interaction is mostly
involved. In this limit the e-ph interaction is strongly
suppressed, particularly in the low-doping regime, where the
effects of the strong e-e interaction play a major role.  A
different behaviour is expected in other quantities where large
frequencies are more relevant. This is in agreement with recent
calculations \cite{zeyher} performed in a single-band Hubbard model
with an on-site "ionic" e-ph coupling, which  show, indeed, that
the Eliashberg function $\alpha^2  F(\omega)$ determining
superconductivity, is much less reduced than the analogous quantity
determining the transport properties.

These remarks are obviously relevant in any dynamical analysis of
the pair formation in strongly interacting systems like, e.g., high
temperature superconducting cuprates, fullerenes
 or ${\rm{Ba_{1-x}K_xBiO_3}}$. In particular, our work
 calls for a critical reanalysis of effective potential models in
the Eliashberg approach of pairing \cite{carbotte} \cite{sham},
when  a strong e-e interaction is present in the systems. Most of
the work in this field has, in fact, been carried out using model
potentials, where the momenta are averaged on the Fermi surface.
This leads to a trivial momentum dependence (likely overestimating
the role of large momenta), which misses the  peculiar
strong momentum vs frequency dependence revealed in the interacting
systems.

On the other hand our general analysis  showed that
the e-ph coupling  can drive the strongly correlated system to an
instability. This was specifically shown for the three-band Hubbard
model in the $U\to \infty$ limit, where more specific features
arise. In particular it was evidenced that the Fermi-liquid system
has diverging long-wavelength density fluctuations in some regions
of the $(\varepsilon_p^0-\varepsilon_d^0)/t_{pd}$ vs. $\delta$ phase
diagram, even with a reasonably small e-ph coupling. This is due to
the phonon induced attraction leading to the violation of the
Landau stability  criterion $F_0^s > -1$. In analogy with other
cases \cite{GRCDK1}-\cite{RCGBK} this appears as an overdamping of
the zero-sound mode. Moreover it was found that superconducting
pairing both in the s- and d-wave channels occurs close to the
instability region. It is worth stressing that the detected Cooper
instabilities occur in the presence of an {\rm {infinite\/}} local
repulsion $U$.

These findings are of obvious theoretical relevance in the context
of  high temperature superconductivity. The general considerations
of Section II show that one is not  allowed to draw any
naive conclusion on the extension of the superconducting region on
the basis of a static analysis, which neglects the presence
of the e-e interaction in the e-ph coupling, once strong
correlations are present. However, it should be noticed that the
results of Section II indicate that, if a system displays Cooper
phenomenon in a static ($\omega =0$) analysis including vertex
correction, the pairing can only be favoured  by the extension of
the analysis to finite frequencies. We also like to point out
that phonon corrections
to the e-ph vertex, which are usually neglected according to the
Migdal theorem \cite{migdal}, have recently been considered
in Ref.\cite{pietronero}. This analysis showed that these
corrections, which are not negligible if the bandwidth
vs phonon frequency ratio is sizable, tend to suppress the
vertex in the large $v_Fq/\omega$ region, whereas they tend
to enhance the e-ph coupling when $v_Fq/\omega$ is small. These
corrections tend therefore to cooperate with the strong
interaction effects discussed in the present paper.

Our results suggest  that, if
high temperature superconductivity is driven by e-ph coupling, this
is possibly due to the fact that the system is close to a PS
 instability, where the argument of the vertex suppression does
not apply. If this is the case, superconductivity itself would
 possibly provide the system a stabilizing mechanism
with respect to superconductivity \cite{CCCDGR}.

By increasing the bare e-ph coupling, one could indeed get
a sizable attractive $\lambda_s(\omega=0)$ at small and intermediate
doping if, for instance, the e-ph mediated
attraction is evaluated at order $\gamma^2$ (a similar
calculation has been carried out in Ref.\cite{tesanovic}
for the single-band infinite-U Hubbard model with a covalent
(intersite) e-ph coupling).
However, at the same time, $F_0^s$ (which is less affected
by e-e vertex corrections), evaluated in the normal state,
should be largely negative ($F_0^s \ll -1$)  and one should
worry about the stability of the system. The straightforward
approach would be to allow for pairing  and then evaluate
${\partial n \over \partial \mu}$ and then check
for thermodynamic stability. This requires the evaluation of
e-e screening effects in the superconducting state, where
hopefully the (partial or full) opening of the gap will
stabilize the system. In this paper we have kept the aptitude
to evaluate the dimensionless couplings $\lambda_l$ in the stable
region where $F_0^s>-1$, leaving to a future work the analysis
of the region with $F_0^s<-1$ in the normal phase.

The generality of the arguments presented in Section II and the
qualitative independence of the results obtained in the framework
of the three-band Hubbard model by varying the e-ph couplings
$G_p$ and $G_d$ and other parameters of the model (like, e.g.,
the phonon frequency $\omega_0$ or the oxygen-oxygen hybridization
$t_{pp}$), witness that our results are rather robust and should
also depend little  on the particular choice of the phonon
considered in the model. The choice taken in the model
(\ref{ham1}) was dictated by simplicity, but we do not expect
qualitative differences in the physics of more realistic models.
Besides
superconductivity, our results provide a possible explanation for
other general features of the superconducting cuprates as well. In
particular  our results concerning the presence of incommensurate
CDW would explain the formation of the superstructures, which are
present in many copper oxides, if this phenomenon has an electronic
origin.

A relevant issue related to the work presented  here concerns the
possible formation of polarons in the three-band Hubbard model.
The treatment of this long standing problem is beyond the scope of
the paper, but it is worth emphasizing that the band narrowing
occurring at low doping due to strong e-e repulsion could ease the
polaron formation.

Some insight on this can be gained from the analysis of Monte
Carlo calculations \cite{deraedt} showing that a polaronic regime
may take place in a single-band model with one single electron
coupled via a short-range interaction $g$
to optical phonons. The polaronic regime occurs as soon as the
e-ph coupling $g$ exceeds a critical value $g_c$. The numerical
analysis shows that the critical value depends on both the
electronic bandwidth ($\sim t$ in Ref.\cite{deraedt}) and the
phonon frequency ($\omega_0$), and it can be represented by the
condition  $g_c^2/(t \omega_0)=\lambda_c \simeq 1$ in the
adiabatic, $t \gg \omega_0$, regime or by the condition
$g_c/\omega_0=\alpha_c \simeq 1$ in the antiadiabatic, $t \ll
\omega_0$, regime.

 The  numerical Monte-Carlo analysis \cite{deraedt}
 only concerns the case of one single electron in a
lattice, whereas the presence of other interacting electrons may
modify the above picture. As discussed in Section II, the high
sensitivity of the vertex corrections as a function of frequency
and momentum renders the above conditions for polaron formation
rather ambiguous, since it is not clear which frequency vs momentum
regions dominate the e-e screening processes dressing the e-ph
coupling $g$ (cf. Section II). A quick inspection to the  $\langle
c^{\dag}_i c_i (a_i+a^{\dag}_i) \rangle$ correlation function shows
that a rather natural guess is that the relevant frequencies are
smaller than the phonon frequency $\omega_0$. Then, if  the
quasiparticle bandwidth $t^*$ is larger than $\omega_0$, the vertex
corrections to the phonon propagator are largely to be considered
in the large momentum ($v_Fq /\omega >1$ and $v_Fq \approx t^*$)
and are large, thus leading to a suppression of the e-ph
correlation. This would disfavor the establishment of a polaronic
regime. On the other hand, as soon as the effective fermion
bandwidth $t^*$ decreases (e.g. approaching the zero-doping
insulating phase) the vertex corrections are in the high frequency
region ($t^*\approx v_Fq < \omega $) and become small, so that the
effective e-e coupling is given by
$\lambda = g^2/(\omega t^*)$, which grows large. This increase
likely  leads to polaron formation.  This considerations suggest
that, if the e-e correlations do not  lead to a fermion effective
bandwidth smaller that the typical phonon frequencies, polaron are
more difficult than in the non interacting system. On the contrary,
if the closeness to an insulating phase reduces the effective
bandwidth determining  non-adiabatic conditions for the
quasiparticles, then the polaronic regime is made easier:  the e-e
interaction itself reduces the electronic kinetic energy thus
favoring the gain in lattice deformation energy related to the
polaron formation. Notice that the
lattice-driven instability of the Fermi liquid, which we
showed to occur in the three-band Hubbard model
for not too large e-ph coupling, does
not appear to be related to the formation of polarons in the
system. In fact the instability occurs
at  small $q$'s (specifically at $q=0$), while the formation of local
polarons involves all $q$s. Moreover the polaron instability is
 associated to
the rapid softening and hardening of the  phonon mode, while our
dynamical analysis has shown that the instability is a consequence
of an overdamping of the zero-sound, the phonon remaining massive.

A last remark can be made concerning the possible formation of
bipolarons. In the limit of weak e-e repulsion $(U\simeq 0)$,
and in the antiadiabatic regime, a large enough phonon-mediated
attraction would lead to a violation of the Landau stability
criterion $F_0^s<-1$
(since $\Gamma_\omega^e\simeq 0$, $\Gamma_\omega^{ph}
<0$ and $\nu^*$ is large) signaling the instability of the Fermi
liquid. This criterion, instead, is not violated in the low-doping
limit of the infinite-$U$ three-band Hubbard model (at least in the
more physical case with  a finite oxygen-oxygen hybridization). This
indicates that, even when the antiadiabatic conditions are realized
(close to the insulating phase), the large repulsion is able to
stabilize the system against bipolaron formation: if polarons
are formed, the  low frequency analysis indicates that they are
stable against bipolaron formation. Of course this does not exclude
the possibility af a dynamical binding of the polarons.

A clear confirmation of our results concerning the presence of
a lattice-driven instability comes from the numerical exact
diagonalization of the three-band Hubbard model  with e-ph coupling
in a small cluster  \cite{lorenzana}. This calculations show
that a CDW occurs at finite doping (one hole in a cluster of four
Cu atoms with surrounding oxygen) if the e-ph coupling does
not exceeds a critical value. Above this value the system
enters, instead, a small-polaron regime. It is quite appealing
to associate the CDW in the small cluster to the unstable region in
the infinite system considered in this paper. Of course,
in the  case investigated in the present paper by continuously
increasing the doping, the instability first occurs at zero
wavevectors, but this  behaviour could hardly be detected in a
finite small cluster, where there are few available  momenta and
where the doping cannot be varied smoothly (the addition of a
single hole already produces a doping around 0.25).


{\it Acknowledgements}---
The authors acknowledge
interesting discussions with Prof. C. Di Castro
and Dr. R. Raimondi. The authors were supported
by the European Economic Community under contract no. SC1$^*$
0222-C(EDB).

\vfill \eject \newpage

{\bf {\centerline{FIGURE CAPTIONS}}}

FIG. 1: Phase diagram
$(\varepsilon_p^0-\varepsilon_d^0)/t_{pd}$ versus positive doping
$\delta$ for $t_{pp}=0$, $G_p=0.15t_{pd}$, $G_d=0.1t_{pd}$, and
$\omega_0=0.02t_{pd}$.

FIG. 2: Phase diagram
$(\varepsilon_p^0-\varepsilon_d^0)/t_{pd}$  versus positive doping
$\delta$ for $t_{pp}=0.2t_{pd}$,
$G_p=0.15t_{pd}$, $G_d=0.1t_{pd}$, and $\omega_0=0.02t_{pd}$.

FIG. 3: Effective static quasiparticle scattering amplitude versus
 transferred momentum well outside
($\delta=0.2$, dotted line), slightly
outside ($\delta=0.225$, continuous line) and
well inside ($\delta=0.3$, dashed line) the unstable
region. The parameter are as in Fig. 2 with
$\varepsilon_p^0-\varepsilon_d^0=4.9t_{pd}$.
The instability occurs at $\delta_c=0.23$. The scattered
quasiparticles are on the Fermi surface and $q$ is in the $y$
direction, ${\bf q}=(0,q)$.

FIG. 4: a) Diagrammatic representation of the
renormalized quasiparticle-phonon vertex.
The electronically screened vertex is reported in b), where
the open dot is the bare phonon-quasiparticle vertex and
the thick dashed line is the resummed slave-boson propagator
diagrammatically represented in c). No phonon propagators are
included in the resummation c).

FIG. 5: Interaction-renormalized quasiparticle-phonon vertex at
fixed low ($q=0.06$, continuous line) and large momenta ($q=2.20$,
dashed line) as a function of
the transferred Matsubara frequency $\omega$.
The parameters are as in Fig.3 with $\delta=0.225$.
The scattered quasiparticles are on the Fermi surface and
$q$ is in the $y$ direction, ${\bf q}=(0,q)$. The inset shows an
enlargement of the low frequency region. The quasiparticle density
of states is $\nu^*\simeq 13/t_{pd}$ per cell.
The vertical lines indicate the phonon frequency
and half the quasiparticle bandwidth.

FIG. 6: Effective dynamic full-phonon-mediated scattering
amplitude  at fixed low ($q=0.06$, continuous line) and large
momenta ($q=2.20$, dashed line)
versus transferred Matsubara frequency. The parameters
are as in Fig.3. The calculations are performed well outside
($\delta=0.2$, diamonds) and slightly outside ($\delta=0.225$,
crosses) the unstable region.  The scattered quasiparticles are on
the Fermi surface and $q$ is in the $y$ direction, ${\bf q}=(0,q)$.
The quasiparticle density of
states is $\nu^*\simeq 13/t_{pd}$ per cell.

FIG. 7: (a) Effective dynamic quasiparticle total (continuous line),
purely electronic repulsive (dashed line) and full-phonon
(dotted line) scattering amplitudes in the Cooper channel
at fixed low momentum ($q=0.06$) as a
function of the  transferred Matsubara frequency.
The parameters are as in Fig.3. The calculations
are performed slightly outside ($\delta=0.225$)  the unstable region.
The scattered quasiparticles are
on the Fermi surface and $q$ is in the $y$ direction, ${\bf
q}=(0,q)$. The quasiparticle density
of states is $\nu^*\simeq 13/t_{pd}$ per cell.
 The vertical lines indicate the phonon frequency
and half the quasiparticle bandwidth.
(b) Same as in (a) but at large momentum ($q=2.20$).

FIG. 8: (a) Effective dynamic quasiparticle total (continuous line),
purely electronic repulsive (dashed line) and full-phonon
(dotted line) scattering amplitudes in the Cooper channel
at fixed low ($q=0.07$) versus
transferred Matsubara frequency. The calculations are performed
at $\delta=0.18$, close to  the unstable region, occurring
at $\delta_c=0.185$. The phonon frequency is
$\omega=0.08t_{pd}$, $G_p=0.3t_{pd}$ and $G_{d}=0.1t_{pd}$.
The scattered quasiparticles are
on the Fermi surface and ${\bf
q}=(0.053,0.03)$ ($|q|=0.06$). The quasiparticle density
of states is $\nu^*\simeq 9/t_{pd}$ per cell.
The vertical lines indicate the phonon frequency
and half the quasiparticle bandwidth.
(b) Same as in (a) but with large momenta ${\bf
q}=(0,2.25)$.

FIG. 9: Effective dynamic quasiparticle total (continuous line),
purely electronic repulsive (dashed line) and full-phonon
(dotted line) scattering amplitudes in the Cooper channel
at fixed low momentum ($q=0.083$) versus
transferred Matsubara frequency. The phonon frequency is
$\omega=0.02t_{pd}$, $G_p=0.15t_{pd}$ and $G_{d}=0.1t_{pd}$ and
the calculations are performed
at $\delta=0.2$, far from the unstable region, occurring
at $\delta_c=0.225$.
The scattered quasiparticles are
on the Fermi surface and ${\bf
q}=(0.078,0.028)$ ($|q|=0.083$). The quasiparticle density
of states is $\nu^*\simeq 9/t_{pd}$ per cell.
The vertical lines indicate the phonon frequency
and half the quasiparticle bandwidth.

\vfill \eject \newpage
{\bf {\centerline{TABLE CAPTIONS}}}

TABLE 1: $s_1$- and $d_1$-wave coupling constants for various doping
and for the model with $\varepsilon_p^0-\varepsilon_d^0=3.3t_{pd}$,
$t_{pp}=0$,
$G_p=0.15t_{pd}$, $G_d=0.1t_{pd}$, and $\omega_0=0.02t_{pd}$.
The instability line is at $\delta_c=0.21$.

TABLE 2: $s_1$-, $s_2$-, $d_1$- and $d_2$-wave coupling constants
for various doping and for the model with
$\varepsilon_p^0-\varepsilon_d^0=4.9t_{pd}$,
$t_{pp}=0.2t_{pd}$,
$G_p=0.15t_{pd}$, $G_d=0.1t_{pd}$, and $\omega_0=0.02t_{pd}$.
The instability line is at $\delta_c=0.23$.

\vfill \eject \newpage

\begin{table}
\centerline{TABLE 1}
  \begin{tabular}{|c|c|c|c|c|}
   \hline
        $\delta$ & 0.15   & 0.20 & 0.208 & 0.22  \\
   \hline
   $\lambda_{s_1}$   & -0.75  & -0.5 & 0.43  & -0.45 \\
   $\lambda_{d_1}$   & -0.044 & 0.2  & 0.35  &       \\
   \hline
   \end{tabular}
\end{table}

\vskip 2truecm

\centerline{TABLE 2}
\begin{table}
  \begin{tabular}{|c|c|c|c|}
   \hline
        $\delta$ & 0.15   & 0.20 & 0.229 \\
   \hline
   $\lambda_{s_1}$   & -0.58   & -0.48 & 1.2  \\
   $\lambda_{s_2}$   & -0.55   & -0.65 & 1.5  \\
   $\lambda_{d_1}$   & -0.063   & 0.017 & 2.1  \\
   $\lambda_{d_2}$   & -0.021   & 0.012 & 1.8  \\
   \hline
   \end{tabular}
\end{table}


\begin{references}
\bibitem{notamax} We recall that the PS is signalled by the
  occurrence of a diverging compressibility enclosing a region of
densities with negative compressibility to be eliminated by a
Maxwell construction.
\bibitem{notagam} This case (which differs from the intersite e-ph
    coupling considered in Ref.\cite{kim}) will allow us to draw
    general conclusions using simple Ward identities,
   Eqs.(\ref{wi1}) and
   (\ref{wi2}) below. This general conclusion is easily extended
    to acustic phonons as well, while the
    extension of our Ward identity analysis to include
    intersite e-ph coupling is presently in progress.
\bibitem{notafl}These conditions are not very restrictive and are
    realized, e.g., in the important case of strongly interacting
    fermions close to a metal-Mott-Hubbard or
    Charge-Transfer insulator transition.
\bibitem{notag}Notice that, within this model, the e-ph couplings
   $G_p$ and $G_d$ do not coincide with the e-ph coupling $\gamma$
   introduced in Section II. In the particular case when
   $G_d=G_p=g$ the correspondence is given by the relation
   $\gamma^2=g^2/\omega_0$.
\bibitem{notaeqil}A different possible approach could also be viable,
   where the phonons are coupled with the density and not with its
fluctuations only. Then the ion equilibrium positions would  depend
on the tight-binding parameters as well as the doping and should be
self-consistently determined.
This latter scheme should be adopted to apply
the Maxwell construction needed to determine the PS region, from the
analysis of the chemical potential $\mu$ versus the filling $n$.
Our simplifying approach assumes
that the atomic level energies already include the shift due to the
non-zero average of the phonon field. We checked that the
two approaches do not give very different results even from the
quantitative point of view.
\bibitem{notamit} The infinite quasiparticle mass is an
outcome of the large-$N$ analysis. At higher order one expects
the narrowing of the bandwidth to be limited by the copper
superexchange J \cite{GKM}, \cite{horsch}.
In the cuprates this could limit the mass
enhancement to $m^*/m \simeq 1.5 t/J\simeq 6-8$ \cite{horsch}
in the small doping limit.
A larger mass enhancement could be expected
in systems with smaller $J/t$.
A large ratio $m^*/m$ is also obtained in
heavy-fermion systems. In the following the limit $m^*/m \to \infty$
should always be intended with the above provisions.
\bibitem{notafix} The static limit is associated to a differentiation
of $G^{-1}$ at fixed chemical potential. Taking the
total derivative of $E_k$ one has to differentiate
with respect to the self-consistency parameters, which, in turn,
depend on $\varepsilon_\alpha^0$ via the self-consistency equations.
These latter equations contain Fermi functions which have
an implicit dependence on
$\varepsilon_\alpha^0$ via the self-consistency
parameters entering $E_k$. Thus, working at fixed chemical potential
amounts to consider
the (total) derivative of the Fermi functions with respect
to $E_k$. On the contrary, this derivative should not be considered
while performing the calculation in the dynamical limit, which
correspond to the fixed doping case.
\bibitem{notahub} Our result is in agreement with the
   one obtained in Ref.\cite{zeyher} for
   the single-band infinite-U Hubbard
   model. However, in that case the effect of the static vertex at
low doping is much less dramatic since this vertex is proportional
to $\left(1+2|\tilde{\mu}|\nu^* \right)^{-1}$, where
$\tilde{\mu}=-2t(\cos (k_{Fx})+\cos (k_{Fy}))$ vanishes
when $\delta \to 0$.
This latter effect simply arises
from the specific form of the Fermi
surfaces in the single-band Hubbard
model, where the constant energy
surfaces are determined by the condition
$\cos (k_{x})+\cos (k_{y})={\rm{constant}}$. Then,
at $\delta=0$, the Fermi energy reduces to a square
where the constant is zero.
This peculiarity of the single-band Hubbard
model is not present once next nearest neighbor
hopping terms are considered,
which modify the shape of the Fermi surface. In this latter case
$\tilde{\mu}$ no longer vanishes at half-filling
and the presence of a large quasiparticle density of
states $\nu^*$ in the denominator should
lead to a large suppression of the e-ph interaction.
\bibitem{notals} We notice, for instance, that the attractive $\lambda_s$ at
   $\delta=0.208$ in Table 1 is still rather small despite the closeness to
   the instability ($\delta_c=0.210$), where $\lambda_s$ is expected
to diverge. This shows that to get a large $\lambda_s$
requires a quite fine tuning of the model parameters.
\bibitem{notakim} Notice that,  according to the analysis
  of Ref.\cite{kim}, the reduction of the e-ph vertex, which
  we obtain in the small frequency limit, also occurs for the
covalent e-ph coupling. In this latter case the e-ph
coupling, being proportional to the hopping $t_{pd}$,
is reduced by a factor $r_0^2$ already at mean field
level; however, a detailed dynamical analysis of
this reduction is still lacking.
\bibitem{notaopt} In the case of optical phonons, which contribute
negligibly to the quasiparticle scattering in the zero temperature limit,
the depression will concern their contribution at
finite temperature.
\end{references}
\end{document}